\documentclass[superscriptaddress,twocolumn,amsmath,amssymb]{revtex4}

\usepackage[dvips]{graphicx}
\usepackage{epsf}
\usepackage{epsfig}
\usepackage{latexsym}
\usepackage{amssymb}
\usepackage{amsfonts,amsbsy}
\usepackage{amsmath}
\usepackage{dcolumn}
\usepackage{bm}
\usepackage{pifont}

\usepackage[latin1]{inputenc}
\usepackage{graphicx}
\usepackage{amsfonts,amsthm,amssymb,amsxtra,amscd,latexsym}

\begin{document}

\title{Disordered driven lattice gases with boundary reservoirs 
and Langmuir kinetics}

\author{Philip Greulich}%
\affiliation{%
Fachrichtung Theoretische Physik, Universit\"at des Saarlandes, Saarbr\"ucken, Germany}

\affiliation{%
Institut  f\"ur Theoretische  Physik, Universit\"at
zu K\"oln D-50937 K\"oln, Germany
}%

\author{Andreas Schadschneider}

\affiliation{%
Institut  f\"ur Theoretische  Physik, Universit\"at 
zu K\"oln D-50937 K\"oln, Germany
}%
\affiliation{%
Interdisziplin\"ares Zentrum f\"ur komplexe Systeme,
University of Bonn, Germany}%

\date{\today}

\begin{abstract}
The asymmetric simple exclusion process with additional Langmuir
kinetics, i.e.\ attachment and detachment in the bulk, is a paradigmatic
model for intracellular transport. Here we study this model
in the presence of randomly distributed inhomogeneities ('defects').
Using Monte Carlo simulations, we find a multitude of coexisting high- 
and low-density domains.
The results are generic for one-dimensional driven diffusive systems
with short-range interactions and can be understood in terms of a 
{\em local} extremal principle for the current {\em profile}. 
This principle is used to determine current profiles and phase diagrams 
as well as statistical properties of ensembles of defect samples.
\end{abstract}

\maketitle

\section{Introduction}
\label{model_intro}

Despite several recent investigations 
\cite{barma1,barma2,DTASEP_krug,jain_DZRP,ShawKL2004,enaud,ChouL2004b,harris_MF,LakatosBC.2006,barma3,PASEP_dis_santen,prot_prod_zia2,DTASEP_num,GreulichS08,GreulichS08a,GrzeschikHS08}  
the influence of sitewise disorder in driven lattice gases is not yet 
fully understood \cite{disorder_stinche}. 

One focus of studies on the influence of disorder and inhomogeneities
was the Asymmetric Simple Exclusion Process (ASEP), especially its
totally asymmetric variant (TASEP). This process is not only believed
to capture the essentials of driven diffusive systems, but its
homogeneous version is exactly solvable
\cite{derrida1,schuetz_dom,BlytheE07}.  The exact solution allows 
to determine the steady state properties analytically without
approximations. These results can then be used as reference system to
study the influence of disorder, inhomogeneities etc.

Here we will study the competition between disorder, realized through
randomly distributed hopping rates associated to the sites in the
TASEP, and {\em Langmuir kinetics}, i.e.\ attachment and detachment
processes in the bulk.  This is not only of theoretical interest due
to the challenges posed by a non-trivial current profile, but also of
direct relevance for the description of intracellular transport. The
model which we will study here has originally been proposed to
describe motor-based transport along microtubules. Although the
microtubules itself are homogeneous, the presence of {\em
  microtubule-associated proteins (MAP)} \cite{CellBook} can create
inhomogeneities which influence the motion of the motors
\cite{mandelkow5}.

In comparison to the ASEP, the current profile in the presence of
Langmuir kinetics is no longer constant. This requires a slightly
different approach since now a "local" point of view becomes necessary.
Our main interest will be in the (local) {\em transport capacity} 
defined in Sec.~\ref{model_sec}.
This important observable is
now also a \emph{local} variable and is of direct relevance for
biological applications.

This paper is organized as follows: In Sec.~\ref{model_sec} we define
the models that are considered here and review some relevant results.
Sec.~\ref{sec-computer} reports results for current and density
profiles obtained by computer simulations.  In Sec.~\ref{sec-theory}
we develop a theoretical framework that helps us to understand the
simulation results and the phase diagram. This theoretical approach is
applied in Sec.~\ref{sec-expval} to compute the probability that a
randomly chosen defect configuration exhibits phase separation.
Finally, Sec.~\ref{sec-summary} gives a summary and conclusions.


\section{Model and Definitions}
\label{model_sec}

We consider driven lattice gases with open boundary conditions and
\emph{Langmuir kinetics} (LK). To be
more specific we focus on the TASEP which is believed to be a
paradigmatic example for this class of dynamic processes.  
Here different extensions considering LK have been proposed, e.g.\ by 
including the diffusion of detached particles \cite{LipowskyKN01,LipowskyK05}.
We will focus on a less detailed model variant, the {\em TASEP/LK}
\cite{pff1,pff2}, which is a TASEP with additional particle creation and 
annihilation in the bulk. 
The TASEP is defined on a lattice of $L$ sites which are numbered from
$j=1$ to $j=L$ beginning at the left. Each site can be occupied by at
most one particle.  The motion of the particles from left to right is
defined by (local) transition rates between adjacent sites.  The
corresponding hopping rates $p_j$ describing the transitions of
particles to their right neighbours are inhomogeneous. We will focus
on a binary distribution with two possible values $p_j=p$ or $p_j=q$
at each site $j$ where $q<p$.  Sites with transition rate $p_j=q$ will
be referred to as \emph{defect sites}, while a site with transition
rate $p_j=p$ is called a \emph{non-defect site}.
In the following we will call a stretch of $l$ consecutive defect
sites a \emph{bottleneck of size $l$}. 

The boundaries of the system are connected to reservoirs so that particles 
can enter at the left end ($j=1$) and leave at the right end ($j=L$). 
The (fixed) densities $\rho_0$ and $\rho_{L+1}$ of the reservoirs
control the effective entry and exit rates, $\alpha:=\rho_0$ and 
$\beta:=1-\rho_{L+1}$, respectively.

Langmuir kinetics is realized by creation and annihalition of
particles in the bulk.  This can be interpreted as particle
exchange with a bulk reservoir and corresponds to attachment and
detachment processes in the biological context.  The corresponding
rates will be considered to be homogeneous, i.e.\ independent of the
position, throughout this paper \footnote{The effects of
  inhomogeneities in the attachment and detachment rates have recently
  been studied in \cite{GrzeschikHS08}.}.

For large system size the investigation is usually simplified by performing
a continuum limit. Since crucial properties, like the bottleneck
lengths in a disordered system, might depend on the system size we
have to specify this limit more carefully. We define a \emph{weak continuum
  limit} where terms of ${\mathcal O}(1/L)$ are neglected while terms
of ${\mathcal O}(1/\ln L)$ are kept, and a \emph{strong continuum
  limit} where we even neglect terms of ${\mathcal O}(1/\ln L)$. In
the following we restrict ourselves to systems where the local
creation and annihilation rates $\omega_a$ and $\omega_d$ are rescaled with
the system size, while the global rates $\Omega_a:=\omega_a \, L$ and
$\Omega_d:=\omega_d \, L$ are kept constant. Hence $\Omega_a$ and
$\Omega_d$ are system parameters while $\omega_a$ and $\omega_d$ are
adjusted to the system size. In particular in the (weak and strong) 
continuum limit, the local rates vanish: $\omega_a,\omega_d\to 0$
for $L\to\infty$. 

In homogeneous regions of these systems there is a unique {\em
  current-density relation (CDR)} $J(\rho)$, usually called
\emph{fundamental diagram} in the context of traffic flow, that
unambiguously gives the current for a given particle density
$\rho=\langle \tau_j \rangle$ on any site \cite{popkov1}, where $\tau_j=0,1$ is the
occupation number of site $j$. 
The CDR of the TASEP has a single maximum. Later, when 
we will also consider more general driven lattice gases, we 
will always assume that their CDR also has a single maximum. 
The maximum is at the point $\rho_M$ and takes the value 
$J_M=J(\rho_M)$.  In this case for a given current
$J$, two possible values for the density, the \emph{high density
  value} $\rho_H(J)>\rho_M$ and the \emph{low density value}
$\rho_L(J)<\rho_M$ exist.

For these systems, the non-conservation of particles can be expressed
by a source term in the equation of continuity of the stationary
state \footnote{Actually $s(\rho)$ can be defined this way.}:
\begin{equation}
\label{continuity_equ}
J_j-J_{j-1}=s(\rho) \, 
\end{equation}
where $J_j$ is the current through the bond between sites $j$ and
$j+1$.  The attachment of particles is assumed to be inhibited by
particles occupying sites, so we assume $s(\rho)$ to be a globally
decreasing function. In fact one can construct models with attractive
interactions where $s(\rho)$ is an increasing function. However, those
systems might exhibit non-ergodic behaviour \cite{schuetz_nonergodic}
that we do not consider here.  Since $\omega_a,\omega_d\to 0$ in the
continuum limit, we also have $s(\rho)\to 0$ in this limit. Hence
locally the current is \emph{almost constant} for large systems and
the CDR is the same as in the corresponding system without LK
\cite{popkov1,diplom}.

The time evolution per time interval $\Delta t$ of the TASEP/LK can
be written in terms of transition rules:

\noindent For $1<j<L$:
\begin{equation}
\begin{array}{lll}
\mbox{Hopping}: & 10\to 01 & \quad\mbox{with probability } p_j\Delta t \\
\mbox{Attachment}: & 0\to 1 & \quad\mbox{with probability }\omega_a \Delta t\\
\mbox{Detachment}: & 1\to 0 & \quad\mbox{with probability } \omega_d \Delta t
\end{array}
\end{equation}

\noindent for $j=1$:
\begin{equation}
\begin{array}{lll}
\mbox{Hopping}: & 10\to 01 & \quad\mbox{with probability } p_1\Delta t \\
\mbox{Entry}: & 0\to 1 & \quad\mbox{with probability } \alpha \Delta t \\
\mbox{Detachment}: & 1\to 0 & \quad\mbox{with probability } \omega_d \Delta t
\end{array}
\end{equation}

\noindent and for $j=L$:
\begin{equation}
\begin{array}{lll}
\mbox{Attachment}: & 0\to 1 & \quad\mbox{with probability } \omega_a\Delta t \\
\mbox{Exit}: & 1\to 0 & \quad\mbox{with probability } \beta \Delta t
\end{array}
\end{equation}
Other transitions are prohibited. Here ``0'' represents empty and ``1'' 
occupied sites. We can write the time evolution of
the density $\rho_j=\langle \tau_j\rangle$ as
\begin{eqnarray}
  \frac{d\rho_j}{dt}(t) &=& p_{j-1}\langle   \tau_{j-1}(t)(1-\tau_j(t))
     \rangle-p_j\langle \tau_j(t)(1-\tau_{j+1}(t))\rangle \nonumber \\ 
  &+& \omega_a   (1-\rho_j(t))-\omega_d \rho_j(t)
\end{eqnarray}
in the bulk and
\begin{eqnarray}
  \frac{d\rho_1}{dt}(t) &=& -p_1\langle \tau_1(t)(1-\tau_2(t))\rangle
  \nonumber \\ 
  &+& \alpha(1-\rho_1(t))-\omega_d \rho_1(t) \\ 
\frac{d\rho_L}{dt}(t) &=& p_{L-1}\langle
  \tau_{L-1}(t)(1-\tau_L(t))\rangle \nonumber \\ 
  &-& \beta \rho_L(t)+\omega_a (1-\rho_L(t))
\end{eqnarray}
at the left and right boundary, respectively.  The parameters
$\alpha,\beta$ correspond to the generic boundary rates
defined before. The source term is
$s(\rho)=\omega_a(1-\rho)-\omega_d\rho$. We call the hopping rates
$p_j$ which are site-dependent properties \emph{intrinsic parameters}
which in the following will be considered as fixed, $p=1$ and $q=0.6$,
if not stated otherwise. In contrast to this we consider the explicit
dependence of the system properties on the \emph{external parameters}
$\alpha,\, \beta,\, \Omega_a$ and $\Omega_d$.  Other driven lattice
gases of the class characterized above can be written in the same way,
while the local parameters might depend on the states in the vicinity
of the sites and additional correlations might occur.  Nonetheless one
can assume that the TASEP/LK is quite universal as a paradigmatic
model \cite{diplom}.

In this work we are especially interested in randomly distributed
defect sites.  Here the \emph{defect density} $\phi$, which is the
probability that a given site is a defect site, serves as an
additional system parameter.  Hence, transition rates are distributed as
\begin{equation}
p_j=\left\lbrace
\begin{array}{ll}
q & \quad \mbox{with probability } \phi \\
p & \quad \mbox{with probability } 1-\phi  
\end{array}
\right. \,\, .
\end{equation}
Defect distributions of this kind are called \emph{disordered}
\footnote{Note that the definition of the term ``disorder'' varies
  throughout literature. In some works also systems with single
  inhomogeneities are called ``disordered'', while we restrict
  ourselves to random defect samples with finite defect density
  $\phi$.}. 
The properties of such systems are not fully
  determined by the defect density $\phi$, but also depend on the
  spatial distribution of the defects. Since these properties can vary from
sample to sample even for fixed system parameters, an investigation of
ensembles of systems (e.g. disorder average) 
rather than single samples is an issue of physical relevance.

In the following sections we will make use of the \emph{particle-hole-symmetry}
exhibited by the TASEP/LK which is invariant under the symmetry operation
\begin{equation}
\label{part-hole-symm}
1 \longleftrightarrow 0,\,\,\, \alpha \longleftrightarrow \beta,\,\,\, j 
\longleftrightarrow L+1-j,\,\,\, \Omega_a \longleftrightarrow \Omega_d \,\, .
\end{equation}
However, the particle-hole-symmetry is not essential for the generic
behaviour, but it allows to reduce the parameter space that needs to
be investigated.

The TASEP/LK with one defect site was already investigated
numerically and analytically in \cite{pff_1def}. Now we want to
generalize these results to arbitrary defect samples. Therefor we
introduce a local quantity, the \emph{transport capacity} $J^*_j$,
which is the site-dependent maximum current that can be achieved by
tuning the external parameters $\alpha,\, \beta,\, \Omega_a$ and
$\Omega_d$ in the continuum limit \footnote{Note that it is important
  that first the external rates are tuned and then the continuum limit
  is taken, since the vanishing of the local bulk influx $s(\rho)$ is
  necessary.}. This quantity will be discussed in detail in section
~\ref{sec-theory}.


\section{Observations by Computer Simulations}
\label{sec-computer}

In this section we summarize some properties of the system that can be
observed with Monte Carlo simulations. Therefor we compare quantities
of the inhomogeneous TASEP/LK with the homogeneous TASEP/LK and the
TASEP with defects. For simulations we used random-sequential update
with fast hopping probability $p=1$. If not specified else, we fix
$q=0.6$ as slow hopping rate. The unit of time is just one timestep so
that probabilities and rates have the same numerical value.

\subsection{Few defects/vanishing fraction of defects}

\begin{figure}[ht]
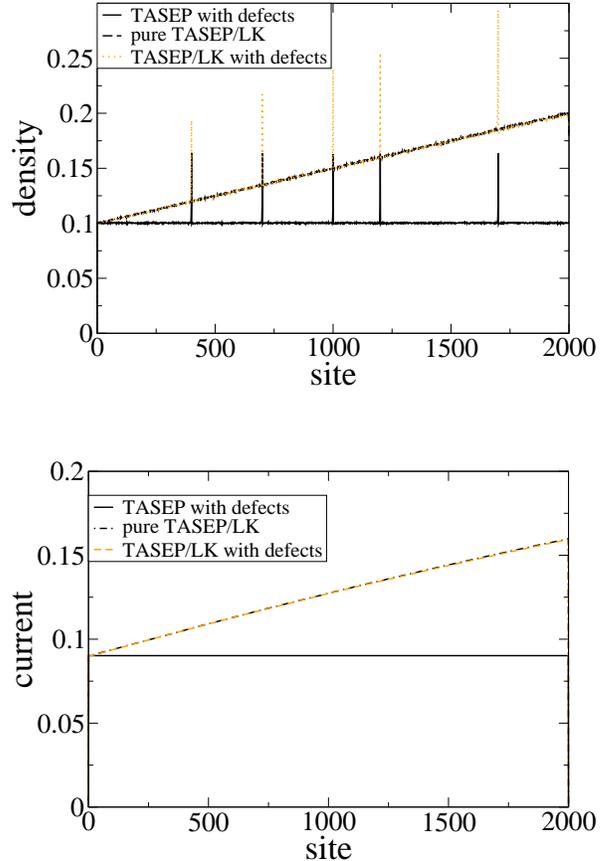

\begin{center}
  \vspace{0.7cm} 
\includegraphics[width=0.9\columnwidth]{DPs_LD_def.eps}\\
\vspace{0.95cm}
\includegraphics[width=0.9\columnwidth]{Jprof_LD_def.eps}
\end{center}
\caption{\label{LD-profs}(Colour online) Comparison of current and 
  density profiles for $\alpha=0.1$ and $\beta=0.9$ (low density
  phase) in the TASEP with defects, homogeneous TASEP/LK and TASEP/LK
  with defects and $\Omega_a=\Omega_d=0.1$.}
\end{figure}

\begin{figure}[ht]
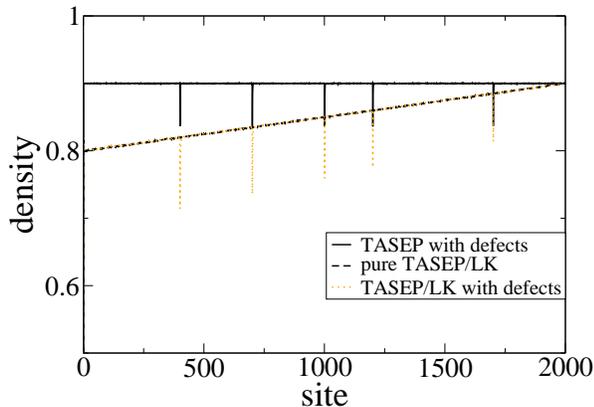

\begin{center}
\vspace{0.5cm}
\includegraphics[width=0.9\columnwidth]{DPs_HD_def.eps}\\
\vspace{0.95cm}
\includegraphics[width=0.9\columnwidth]{Jprof_HD_def.eps}
\end{center}
\caption{\label{HD-profs}(Colour online) Comparison of current and 
  density profiles for $\alpha=0.9$ and $\beta=0.1$ (high density
  phase) in the TASEP with defects, homogeneous TASEP/LK and TASEP/LK
  with defects.}
\end{figure}

\begin{figure}[ht]
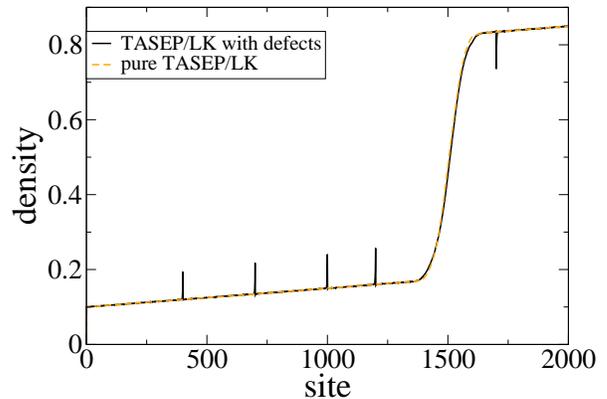

\begin{center}
\vspace{0.5cm}
\includegraphics[width=0.9\columnwidth]{DP_S_def.eps}\\
\vspace{0.95cm}
\includegraphics[width=0.9\columnwidth]{Jprofs_S_def.eps}
\end{center}
\caption{\label{S-profs}(Colour online)   Comparison of current and 
  density profiles for $\alpha=0.1$ and $\beta=0.15$ (high density
  phase) in the homogeneous TASEP/LK and TASEP/LK with defects.}
\end{figure}

Before we consider finite defect densities $\phi>0$ we discuss
systems with a fixed number of defects in the continuum limit ($\phi=0$).
Figs.~\ref{LD-profs}--\ref{S-profs} display the
dependence of the densities and the current on the position in the
system.

Fig.~\ref{LD-profs} shows the density and current profiles of a
TASEP/LK-system with five defects, a homogeneous TASEP/LK-system and a TASEP
with five defects in the low density phase. The density profiles of
inhomogeneous and homogeneous TASEP/LK-systems differ only in the
occurrence of narrow density peaks at the defects, while globally
the density profile is the same. The current profiles
of the homogeneous and inhomogeneous system are identical. In contrast, the
density profile of the TASEP with defects at the same sites shows
density peaks as well, but the current profile (and the density
profile far from the boundaries) is flat. This is due to particle
conservation while the lateral influx of particles allows a spatial
variation of the current profile in the TASEP/LK where particles are
not conserved in the bulk.

Fig.~\ref{HD-profs} shows the corresponding situation for low exit
rate and high entry rate.  Due to particle-hole-symmetry, the results
are analogous to the previous case.  Adopting the terminology of the
homogeneous system, the inhomogeneous TASEP/LK-system can be considered to
be in a high and low density phase, respectively.

Fig.~\ref{S-profs} displays density profiles for
$\alpha\approx\beta$. As in the case above, homogeneous and inhomogeneous
TASEP/LK-systems exhibit the same density profiles, apart from the peaks.
In this case we see a \emph{shock} in the density profile which is 
characteristic for non-particle-conserving dynamics in the bulk and which
cannot be observed in the particle conserving TASEP (except at
$\alpha=\beta$).

\begin{figure}[ht]
\begin{center}
\vspace{0.5cm}
\includegraphics[width=0.9\columnwidth]{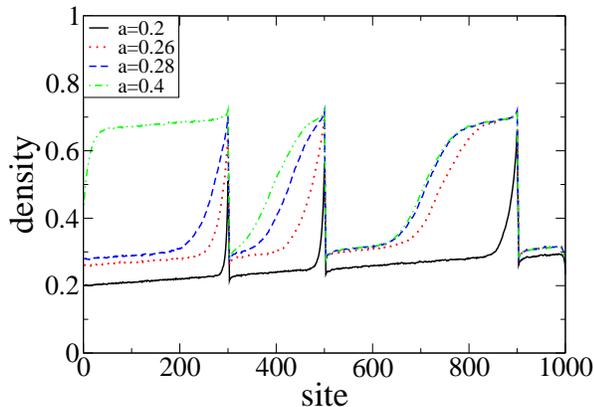}
\end{center}
\caption{\label{phase_sep}(Colour online) Density profiles for 
  increasing values of $\alpha$ and fixed $\beta=0.9$. At a critical
  value $\alpha^*$ a high density region at the most right defect
  occurs (phase separation). For higher $\alpha$ multiple high density 
  regions appear.}
\end{figure}

\begin{figure}[ht]
\begin{center}
  \vspace{0.5cm} 

  \includegraphics[width=0.9\columnwidth]{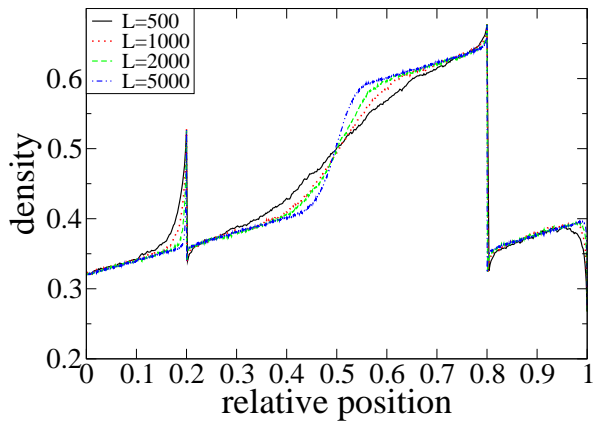}

\end{center}
\caption{\label{finsize_scale1}(Colour online) Density profiles for identical 
  macroscopic parameters $\Omega_a=0.1,\,\, \Omega_d=0.1,\,\,
  \alpha=0.35,\,\, \beta=0.9$ but different system sizes $L$. The left
  boundary of the high density region (shock) becomes steeper with
  increasing system size, indicating a macroscopic regime.
}
\end{figure}

Increasing the entry rate $\alpha$ for fixed and large $\beta$ one
observes a queuing transition in Fig.~\ref{phase_sep}: At a critical
entry rate $\alpha^*$ the peak at the leftmost defect broadens,
forming a high density region. This corresponds to phase separation
and is also observed in the inhomogeneous TASEP at critical boundary
rates.  In the TASEP, however, the high density regime always extends to
the left boundary. In contrast, the inhomogeneous TASEP/LK-system exhibits a
\emph{stationary} shock separating the low and high density region.
Numerical finite-size scaling in Fig.~\ref{finsize_scale1} shows
that the shock is getting sharper with increasing system size.
Thus the high density region extends over a
finite fraction of the system, corresponding to phase separation. In
contrast, the peaks diminish for larger systems indicating that they are
just local phenomena. We can associate this phase separation with a
phase transition at the critical parameter value $\alpha'$.

Increasing $\alpha$ further moves the shock position to the left.
The density profile right of the defect where phase separation
occurred does not change anymore by varying the entry rate. The same is 
true for the \emph{output current} at the right boundary $J_\text{out}=J(L)$.  
At some value of $\alpha$ a second high density region starts to form. Thus in
a system with many defects multiple shocks can occur associated with
alternating domains of high and low density.

Above a critical value $\alpha^*$, where a high density domain extends
to the left boundary, the density profile and the current in the
system is independent of the entry rate. Since this independence also
holds for large $\beta$, we call this a \emph{Meissner phase} in
analogy to superconductors, where the magnetic field in the interior
bulk is independent of exterior fields. This terminology was also used 
for the boundary independent phase in the homogeneous TASEP/LK \cite{pff2}. 
However, one has to note that while
in the homogeneous system there are long-range boundary layers in the density
profile which \emph{do} depend on boundary rates, the Meissner phase
in the disordered system only exhibits short-range boundary layers.
The current profile in fact does not depend on the boundary rates,
both in the homogeneous and the inhomogeneous system.

Due to particle-hole-symmetry all considerations made in this section 
can be transferred to the high density phase by replacing 
$\alpha$ with $\beta$.

\subsection{Finite fraction of defects and disordered systems}
\label{sec_MC_finphi}

\begin{figure}[ht]
\begin{center}
\vspace{0.5cm}
\includegraphics[width=0.9\columnwidth]{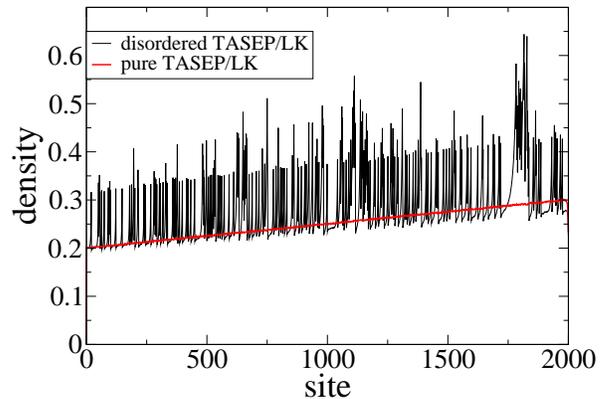}\\
\vspace{0.95cm}
\includegraphics[width=0.9\columnwidth]{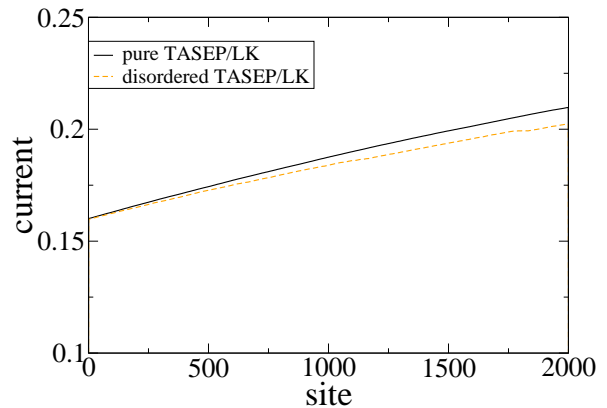}
\end{center}
\caption{\label{LD-profs2}(Colour online) Comparison of current and density 
  profiles for $\alpha=0.1$ and $\beta=0.9$ (low density phase) in the
  disordered TASEP/LK with defect density $\phi=0.1$ and homogeneous TASEP/LK.}
\end{figure}

If the density of defects $\phi$ is finite and the number of defects is of
order of the system size, even a local increase of the density in
the vicinity of the defects has considerable impact on the average density 
due to the large number of defects.
The effect can be observed in Fig.~\ref{LD-profs2} where we have
simulated disordered systems with small but finite defect density  $\phi$
for small $\alpha$ and large $\beta$. In contrast to systems with few
defects, the current profile of the disordered system differs from the
that of the homogeneous system. This is due to the change of the density by
defects, which leads to an altered influx of particles in the bulk by
attachment/detachment. So the gradient of the current profile in the
disordered system is different from the one in the homogeneous system and also
from the system with few defects because in the latter the effect on
the average density is negligible. 

\begin{figure}[ht]
\begin{center}
\vspace{0.5cm}
\includegraphics[width=0.9\columnwidth]{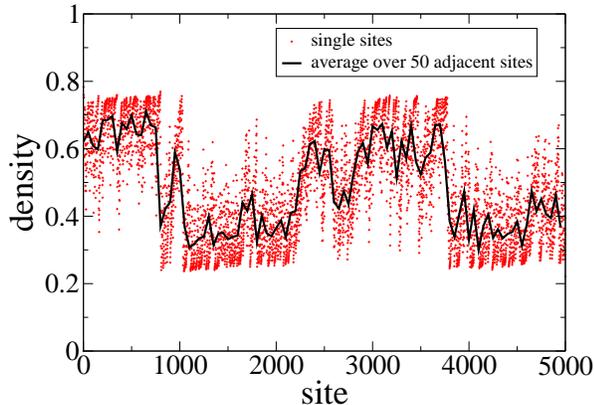}

\end{center}
\caption{\label{phase_sep2} (Colour online) Density profile for $\alpha=0.9$ 
  and $\beta=0.9$ in the disordered TASEP/LK with defect density
  $\phi=0.2$. One observes phase
  separation with alternating high and low density domains. The black
  line displays the density, averaged over 50 adjacent sites.}
\end{figure}

Like in the TASEP/LK with few defects we observe multiple high and 
low density domains for large boundary rates, which is displayed in 
Fig. ~\ref{phase_sep2}. In fact it is harder to distinguish macroscopic
 high and low density regimes in the
disordered case because of the rapid changes of density on a
microscopic scale. We have to simulate rather large systems in order
to identify a macroscopic high(low) density domain by inspection.
In Sec.~\ref{sec-expval} we introduce a numerical method that can 
detect high and low density domains automatically.

\section{Theoretical treatment}
\label{sec-theory}

In this section we develop a theoretical framework for the
observations made by Monte Carlo simulations. We expect that concepts
developed in this section are generic for a larger class of disordered
driven lattice gases that have a single maximum in the current-density
relation and weak induced effective interactions between defects. The
restriction ``weak interaction'' is discussed in detail in
\cite{GreulichS08a}. In addition, we assume that the bulk influx term
$S(\rho)$ is decreasing with increasing density.

First we summarize the properties that distinguish the inhomogeneous
(disordered) TASEP/LK from the TASEP and homogeneous TASEP/LK,
respectively.

\begin{enumerate}
\item In the TASEP/LK the particle number is not conserved in the
  bulk.  Therefore generically the current profile is \emph{not flat}
  and \emph{stationary shocks} can occur in the bulk.  For particle
  conserving systems these are not possible \cite{popkov1,evans_pff}.

\item In the homogeneous TASEP, the current is restricted by the upper
  bound $J^{\rm max}_{\rm hom}=p(1-p)=0.25$ (for hopping rate $p=1$)
  due to the bulk exclusion.  Already a single defect site with lower
  hopping rate $q<p$ reduces this maximum stationary current
  \cite{lebowitz1}. In \cite{pff_1def} it was shown that also in the
  TASEP/LK a single defect site $d$ restricts the current by a value
  $J^*_d:=J^{\rm max}_d<J^{\rm max}_{\rm hom}$ at this site, that
  cannot be exceeded by tuning external parameters. The quantity
  $J^*_d$ is exactly the local \emph{transport capacity} defined in
  Sec.~\ref{model_sec}. However, due to the spatially varying current,
  this effect is only \emph{local} and the maximum value of the
  current $J^{\rm max}_i$ on sites $i$ far away from the defect can be
  larger than $J^*_d$. For completeness, we define $J^*_i=J^{\rm
  max}_{\rm hom}$ on non-defect sites $i$, so the transport capacity
  is peaked on a single site. If the current imposed by the
  boundary rates is larger than the transport capacity of a defect,
  phase separation occurs, exhibiting stationary shocks. In the inhomogeneous
  TASEP no stationary shocks can occur in the bulk, thus the high
  density regime always fills the whole system left of the current
  limiting defect.

\item In systems with only few defects the relation between the
  average density and the current at a given site is the same as in
  the homogeneous system. Thus current profiles are almost the same
  (as long as the maximum current is not exceeded). In disordered
  systems with a finite fraction of defects, however, the
  current-density relation is not the same as in the homogeneous
  system and depends on $q$ and the distribution of defects, since the
  large number of density peaks have influence on the source term
  $s(\rho)$ in (\ref{continuity_equ}) on a macroscopic scale.
  Therefore the current profiles differ from the homogeneous case.
\end{enumerate}

In order to capture these properties, we follow the concept of
\cite{pff_1def} by focusing on the current profiles $J(x)$.

\subsection{The influence of defects: additional initial conditions}

Locally the current profiles are determined by the continuity equation 
(\ref{continuity_equ}). Introducing the continuous variable
$x:=\frac{i-1}{L-1}$, which is the relative position in the system, one
can write $J_{i-1}=J_i-\frac{1}{L}\frac{dJ}{dx}+{\cal O}(1/L^2)$. 
In the stationary state the continuity equation (\ref{continuity_equ}) 
becomes
\begin{equation}
\label{dJdx}
\frac{dJ}{dx} = S(\rho)+{\cal O}(1/L)
\end{equation}
where the global source term $S(\rho)=L\, s(\rho)$ was introduced. 
In the TASEP/LK, for example, we have
$S(\rho)=\Omega_a(1-\rho)-\Omega_d\rho$. In the continuum limit we
neglect terms of $\mathcal{O}(1/L)$ so that (\ref{dJdx}) becomes an
ordinary first order differential equation in the continuous variable
$x$. The system, however, has at least two initial conditions (e.g.
the boundary conditions in the homogeneous case), thus it is
overdetermined. Each initial condition at a point $x_0$ is associated
with one solution of the differential equation (\ref{dJdx})
$J_{x_0}(x)$ for the current and $\rho_{x_0}(x)$ for density,
respectively. We call the mathematical solutions to single initial
conditions $J_{x_0}(x)$ and $\rho_{x_0}(x)$ \emph{local
  current/density profiles}. Physically these solutions are not
necessarily realized.

For the TASEP/LK with a single defect it was shown by Pierobon et al.
\cite{pff_1def} that the finite transport capacity at the defect site,
corresponding to a local upper bound of the current, can be regarded
as an additional condition on the current profile. They argued that
the local solution of (\ref{dJdx}) with the initial condition
$J(x_d)=J^*(x_d)$ becomes relevant if the local current profiles of
the boundary conditions exceed $J^*$ at the defect site. Here we
want to justify this approach and generalize it to a larger class of
driven lattice gases with many defects, including randomly disordered
systems, that meet the restrictions noted earlier in this section.

In \cite{GreulichS08a} it was shown that the maximum current in
particle conserving driven lattice gases with randomly distributed
defects but low defect density depends approximately only on the size
of the longest bottleneck \emph{(Single Bottleneck Approximation,
  SBA)}. This fact, together with the observations made in
\cite{pff_1def}, motivates the generalization of the transport
capacity to driven lattice gases (including TASEP/LK) with many
defects but low defect density, introducing an approximation similiar to SBA.
We call it the \emph{locally independent bottleneck approximation
  (LIBA)}: The transport capacity at a site $x$, $J^*(x)$, is
approximately equal to the maximum current that can be achieved by
tuning the boundary rates in the corresponding system containing {\em only
one} bottleneck at this site
\footnote{In this terminology a non-defect site is also called a
  bottleneck of size 0.}.  Thus $J^*(x)$ can be obtained by
  refering to a single-bottleneck system where all other defects
(except the bottleneck at site $x$) have been removed.

In systems without LK the current is spatially constant and cannot
exceed the minimum of $J^*(x)$ which corresponds to the transport
capacity of the longest bottleneck, since in 
single bottleneck systems the maximum current is equal to the local
transport capacity $J^*(x)$ and decreases with $l$
\cite{GreulichS08,prot_prod_zia2,bn_quasiexact}. In this case the LIBA
reduces to the SBA.

The LIBA neglects the influence of other defects on the transport
capacity at site $x$. Nonetheless, we claim that the influence of
other defects on the transport capacity can be considered as a
perturbation in the same way as it is the case for the SBA in particle
conserving systems \cite{GreulichS08}. Since the local attachment and
detachment rates vanish in the continuum limit, the transport capacity
of a bottleneck should be the same as in the corresponding particle
conserving system. Therefore $J^*(x)$ is independent of $\Omega_a$ and
$\Omega_d$. For the TASEP without LK analytical results are available
\cite{GreulichS08,bn_quasiexact} that can be used to obtain
approximations for the transport capacity. Since the maximal current
in these systams depends only on the bottleneck length $l(x)$
\cite{GreulichS08,prot_prod_zia2} this holds also for the transport
capacity. The concept of a local transport capacity is applicable if
interactions of defects near a bottleneck are not to large and
distances of defects are not too small (i.e.\ low defect density
\footnote{In fact for the disordered TASEP the approximation 
    turns out to be rather robust even for higher defect density.}).

Hence, the transport capacity $J^*(x)$ yields an upper bound for
the current profile,
\begin{equation}
\label{init_cond_1}
J(x) \leq J^*(x) \hspace{5mm} \mbox{ for all $x$,} 
\end{equation}
while the function $J^*(x)$ of course is not continuous. Since on
non-defect sites (which correspond to bottlenecks of size $l=0$) the
transport capacity is $J^*=J^{\rm max}_{\rm hom}$, it is sufficient to
check condition (\ref{init_cond_1}) for defect sites. Their number is
finite in finite systems but can be infinite in the continuum limit
(e.g.\ for disordered systems with finite defect density).

The problem of condition (\ref{init_cond_1}) is that it is given
as an inequality and does not provide initial conditions for
(\ref{dJdx}) on the defect sites. We now want to show that
(\ref{init_cond_1}) is identically fullfilled by a set of initial
conditions
\begin{equation}
\label{init_cond_2}
J(x) = J^*(x) \hspace{5mm} \mbox{ at defect sites $x$} \,\, ,
\end{equation}
if one assumes additionally that the physical local solution at $x$ 
is selected by shock dynamics.

\begin{figure}[ht]
\begin{center}
  \vspace{0.5cm} 
  \includegraphics[width=0.7\columnwidth]{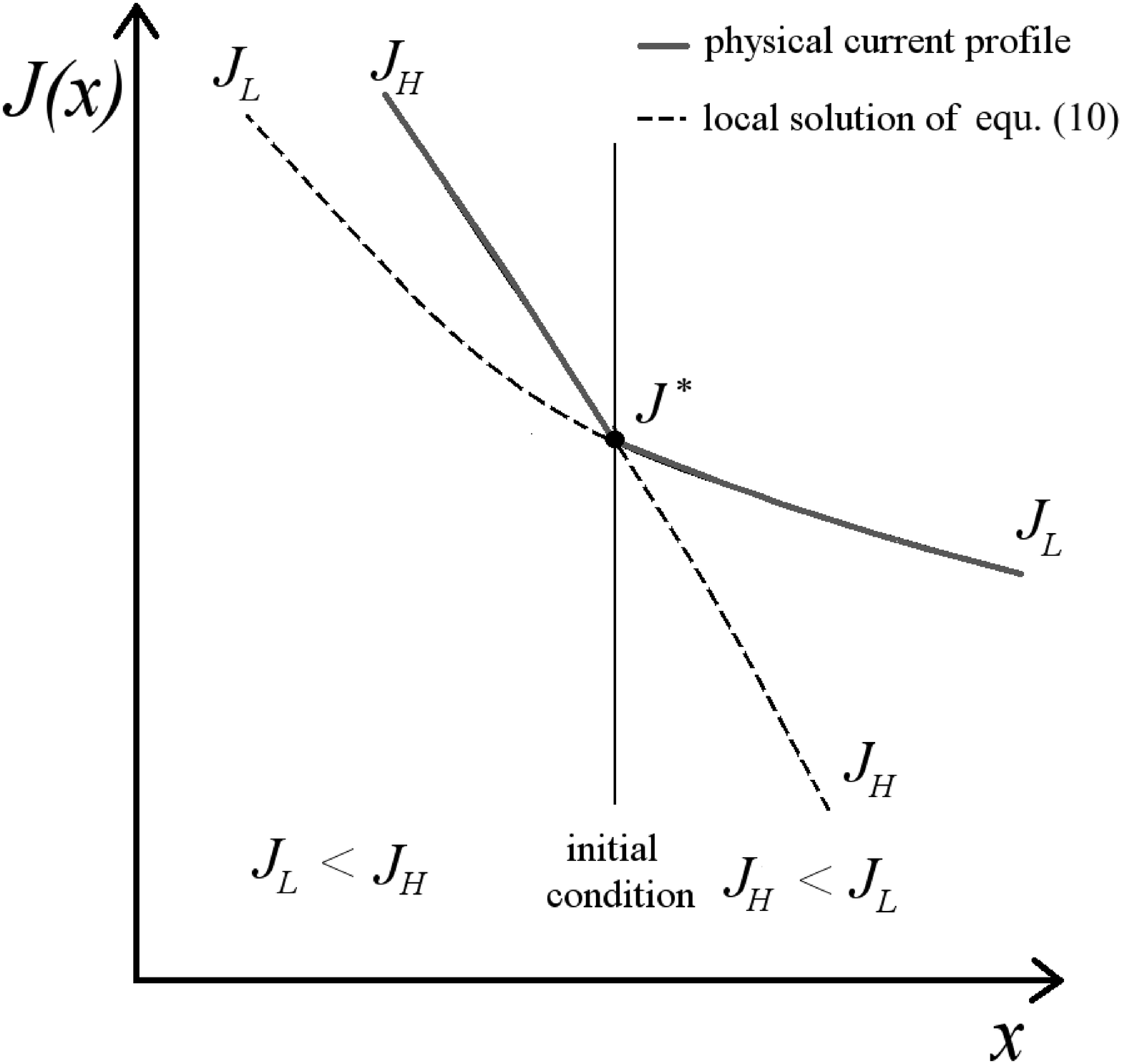}\\
  \includegraphics[width=0.7\columnwidth]{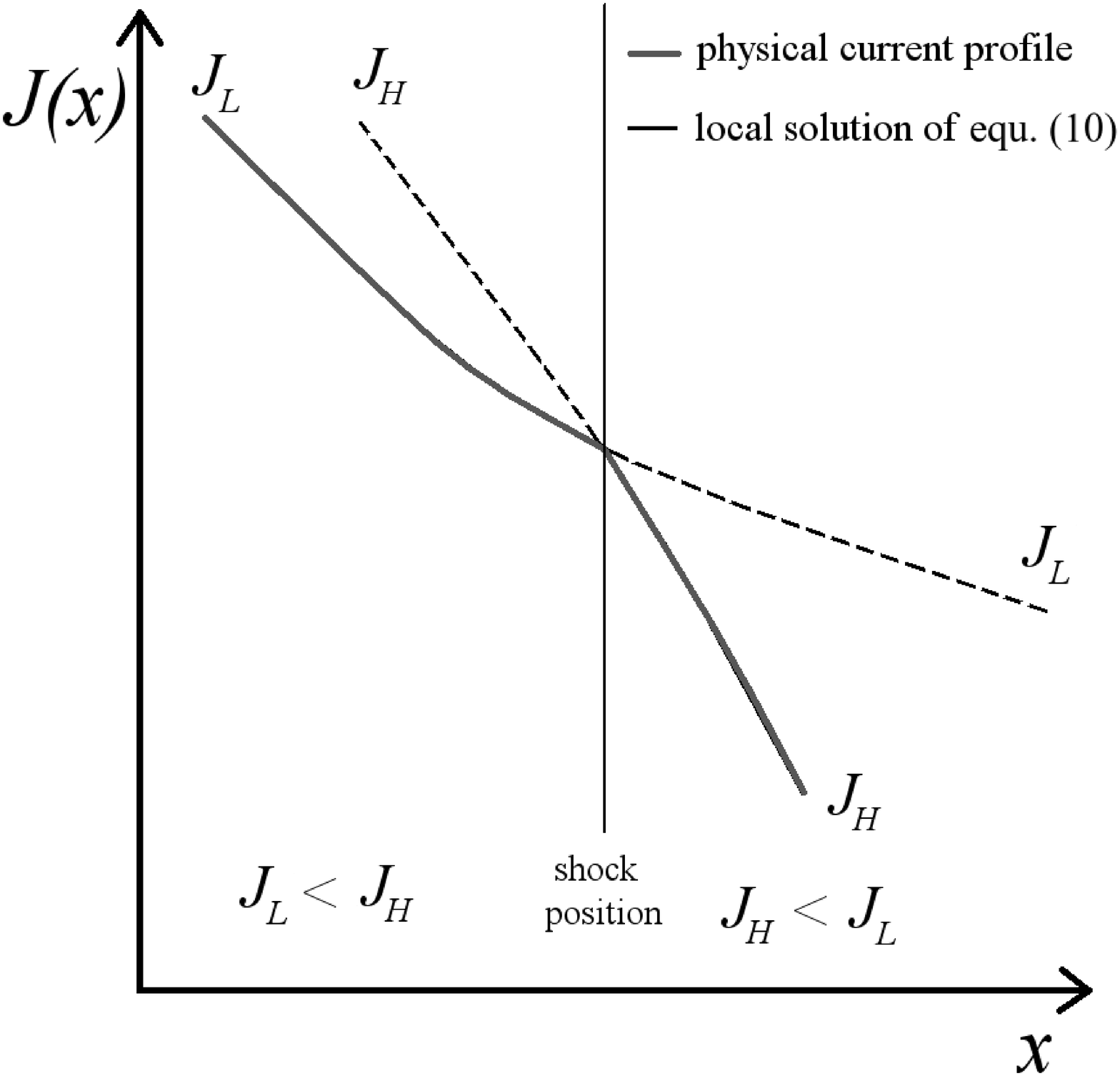}
\end{center}
\caption{\label{initpoint_fig} Top: Local solutions in the vicinity of a point 
  with an initial condition $J^*$. Due to the non-unique inversion of the
  current-density relation, there are two possible solutions. Since
  for a physical solution the direction of the collective velocity
  must point away from this position, only solutions with \emph{maximal
    current} are realized.\\
  Bottom: Intersection point of local solutions of 
  the density profile. The constraint that only upward shocks can
  exist implies that only solutions with \emph{minimal current} are
  physically realized. 
}
\end{figure}

First of all, if we assume the conditions (\ref{init_cond_2}) we see
that, in contrast to the boundary conditions of the system which are
usually given by a fixed density, the initial condition imposed by a
defect provides the possibility of two realizations of the local
density profile.  Given the initial condition $J(x_0)=J^*(x_0)$ at a
point $x_0$, only the current is a fixed initial condition while, due
to the non-unique inversion of the current-density relation (one
maximum!), there are two possible values for the density, $\rho_H$ and
$\rho_L$ (with $\rho_H>\rho_L$), leading to two possible local
solutions of (\ref{dJdx}), a \emph{high density solution} $J_H(x)$ and
a \emph{low density solution} $J_L(x)$:
\begin{equation}
J^* \begin{array}{l}\nearrow \\ \searrow \end{array}
\begin{array}{l}
\rho_H \longrightarrow J_H(x-x_0,J^*) \\ \\
\rho_L \longrightarrow J_L(x-x_0,J^*)
\end{array}
\end{equation}
Taking into account shock dynamics, a constraint on the selection of a
physical solution is given by the collective velocity
\begin{equation}
v_c(x)=J'(\rho(x))
\end{equation}
where $J(\rho)$ is the current density relation and the prime denotes
the derivative with respect to $\rho$ \cite{kolomeisky_shockdyn}.  A
solution can only propagate away from the initial point if the
direction of $v_c$ is pointing away from it, i.e. left of it only
solutions with $v_c<0$ can exist, while right of it solutions must
have $v_c>0$. In a system with a single maximum at density $\rho_m$ in
the CDR, $\frac{dJ}{d\rho}>0$ for $\rho<\rho_m$ and
$\frac{dJ}{d\rho}<0$ for $\rho>\rho_m$, thus left of an initial point,
only the high density solution $J_H$ can be realized, while right of
it only $J_L$ can physically exist. This principle is displayed in
Fig.~\ref{initpoint_fig}, top. Hence, each initial condition at a
point $x_0$ can have its own solutions. We denote these \emph{local
  solutions} by
\begin{equation}
J(x-x_0,J^*)=\left\lbrace\begin{array}{lc}
J_H(x-x_0,J^*) & \mbox{for $x \leq x_0$} \\
J_L(x-x_0,J^*) & \mbox{for $x>x_0$}
\end{array}\right. \,\, .
\end{equation}
Actually the dependence on $J^*$ can easily be obtained by a shift
operation if two functions $\tilde J_L(x)$ and $\tilde J_H(x)$ with
initial conditions $\tilde J_{L}(0)=J^0_L$ and $\tilde J_{H}(0)=J^0_H$, 
where $J^0_L$ and $J^0_H$ are arbitrary chosen values in
the high- and low density branch of the CDR. If the range in both
branches of the CDR includes $J=0$, one can simply choose
$J^0_L=J^0_H=0$ \footnote{Note that this is the case for systems with
  strict exclusion interaction like TASEP and TASEP/LK. If double
  occupancy is possible, the CDR not necessarily vanishes for
  $\rho=1$.}. Since the ODE (\ref{dJdx}) is of first order and does
not explicitly depend on $x$, the high and low density solutions
unambigiously depend on $\rho$ and are monotonic. Thus different local
solutions $J_{L,H}$ can only differ by a shift in the variable $x$. An
arbitrary solution $J_{L,H}(x-x_0,J^*)$ can be obtained by shifting
$\tilde J_{L,H}(x)$ by an amount $\tilde x_{L,H}(J^*)$ so that the
value of the shifted function at $x=0$ is equal to $J^*$. The
functions $\tilde x_{L,H}(J^*)$ are just the inverse functions of the
unique functions $\tilde J_{L,H}(x)$.  Then the local solutions at a
point with initial condition $J^*$ are given as
\begin{equation}
\label{Greensfunc}
J(x-x_0,J^*)=\tilde J(x-x_0-\tilde x(J^*))\,.
\end{equation}
The functions $\tilde J_{L,H}(x)$ and $\tilde x_{L,H}(J)$ can for
example be obtained by numerical solution of (\ref{dJdx}) with initial
conditions $J^0_{L,H}$.

\subsection{Selection of the global current profile}
\label{global_sec}

The physically realized \emph{global} current profile in the steady
state is also determined by shock dynamics
\cite{popkov1,kolomeisky_shockdyn}. Shocks manifest themselves as
discontinuities in the density profiles. If they are stationary they
connect different local steady state solutions of (\ref{dJdx}) to form
a \emph{global solution}. The crucial quantity for this selection is
the shock velocity
\begin{equation}
v_s=\frac{J_+-J_-}{\rho_+-\rho_-}
\end{equation}
that determines the propagation of a discontinuity in a (not
necessarily stationary) density profile. Here $J_+$($\rho_+$) is the
current (density) right of the shock and $J_-$ ($\rho_-$) is the
current (density) left of the shock. In homogeneous driven lattice
gases with a single maximum in the CDR only upward shocks with
$\rho_+>\rho_-$ can exist (see for example
\cite{kolomeisky_shockdyn,shockdyn1}). In \cite{popkov1} this was
generalized to systems with particle creation and annihilation in the
bulk, as long as the local creation and annihilation rates vanish in
the continuum limit, i.e. $s(\rho)\to 0$ for $L\to\infty$. In this
case, the CDR is the same as in the corresponding particle conserving
system.

In inhomogeneous systems there can also be ``downward'' discontinuities at
the defect sites due to the imposed maximum current. However, these
discontinuities usually are not called ``shocks'' since their
dynamics differ. In contrast to shocks they are sharp also in finite
systems, thus there are no fluctuations. Due to the local character of
$v_s$ and $v_c$ we can state that away from defects, where locally the
system is homogeneous, only upward shocks can exist.


Since the source term $s(\rho)$ of (\ref{continuity_equ}) vanishes in
the continuum limit, shocks can only be stationary at intersection
points of a high and a low density solution $J_H(x)$ and $J_L(x)$. So
only at these intersection points a switch of the physical realized
local solution can occur.  Note that local solutions of the same kind
$J_L$ or $J_H$ cannot intersect since the differential equation
(\ref{dJdx}) is of first order. Since $S(\rho)$, which determines the
slope of the current profile, is assumed to be a monotonically
decreasing function in $\rho$, we have $S(\rho_H)<S(\rho_L)$, hence
the gradient of the high density solution $J_H(x)$ is smaller than the
one of the low density solution $J_L(x)$.  Therefore left of an
intersection point, we have $J_L(x)<J_H(x)$, while right of it
$J_H(x)<J_L(x)$. Since $J_L$ is the physical solution left of a shock
and $J_H$ right of it, always the minimal
local solution is the physical one (see Fig.~\ref{initpoint_fig}, bottom). 
We define the minimal envelope of all the local current profiles as the
\emph{capacity field} of the system
\begin{equation}
\mathcal{C}(x):=\min_{x'}\left\lbrace J(x-x',J^*(x')) \right\rbrace
\end{equation}
with defects at the points $x'$. This function does not depend on the
boundary rates. The capacity field is a generalization of the capacity
introduced in \cite{pff_1def}. Note that in general the capacity
field is not identical with the local transport capacity $J^*(x)$
\footnote{For example a single defect at site $x_d$ and maximum
  current $J^*_{1{\rm def}}$ has a peaked local transport capacity
  $J^*(x)=J^*_{1{\rm def}} \delta(x-x_d)$, while the capacity
  $\mathcal{C}(x)$ is an extended function.}. The local transport
capacity can be viewed as the source or ``charge'' of the capacity
field. In this view, the function $\tilde J_{L,H}(x-x_0)$, which 
generates all local current profiles via (\ref{Greensfunc}), can be
called the ``Green's function'' of the capacity field.

Additional conditions on the current profile are given by the boundary
rates so that $\rho(0)=\alpha$ and $\rho(1)=1-\beta$. Of course the
maximum current of the homogeneous system $J^{\rm max}_{\rm hom}$
remains an upper bound also in the inhomogeneous system. The capacity
field together with the boundary conditions can be used to express the
physically realized current profile as
\begin{equation}
\label{J(x)_glob}
J(x)=\min\left[J_{\alpha}(x),J_{\beta}(x),\mathcal{C}(x) \right]
\end{equation} 
This principle is the generalization of the extremal current principle
for the homogeneous TASEP \cite{popkov2}.  It provides a tool to obtain
the global current profile if it is possible to obtain the local
solutions of (\ref{dJdx}) and the local maximum current $J^*(x)$.

Indeed the global current profile given by (\ref{J(x)_glob})
identically fulfills the condition (\ref{init_cond_2}) that the
current must always be lower than the transport capacity.

\begin{figure}[ht]
\begin{center}
\vspace{0.5cm}
\includegraphics[width=0.9\columnwidth]{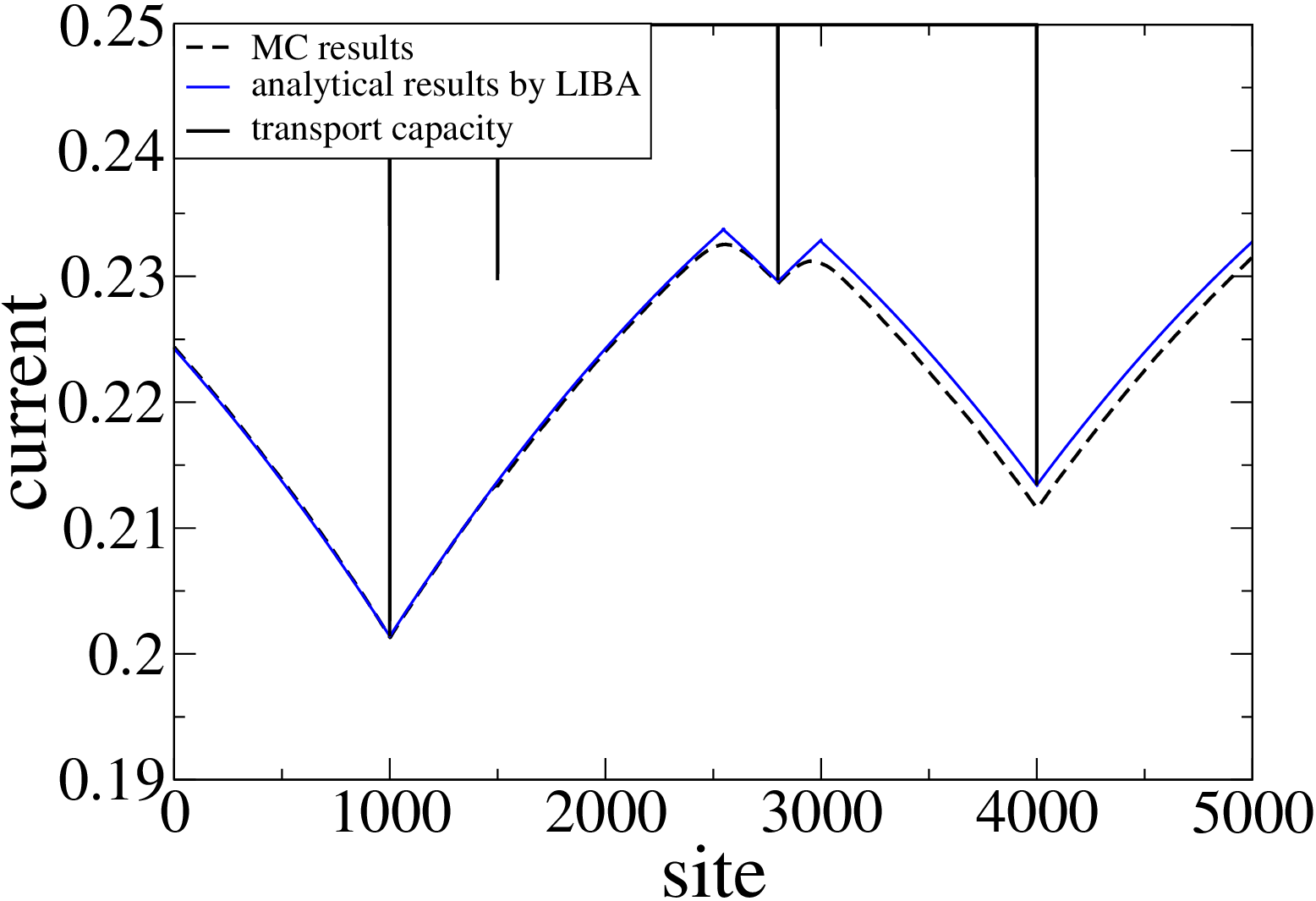}
\end{center}
\begin{center}
\vspace{0.5cm}
\includegraphics[width=0.9\columnwidth]{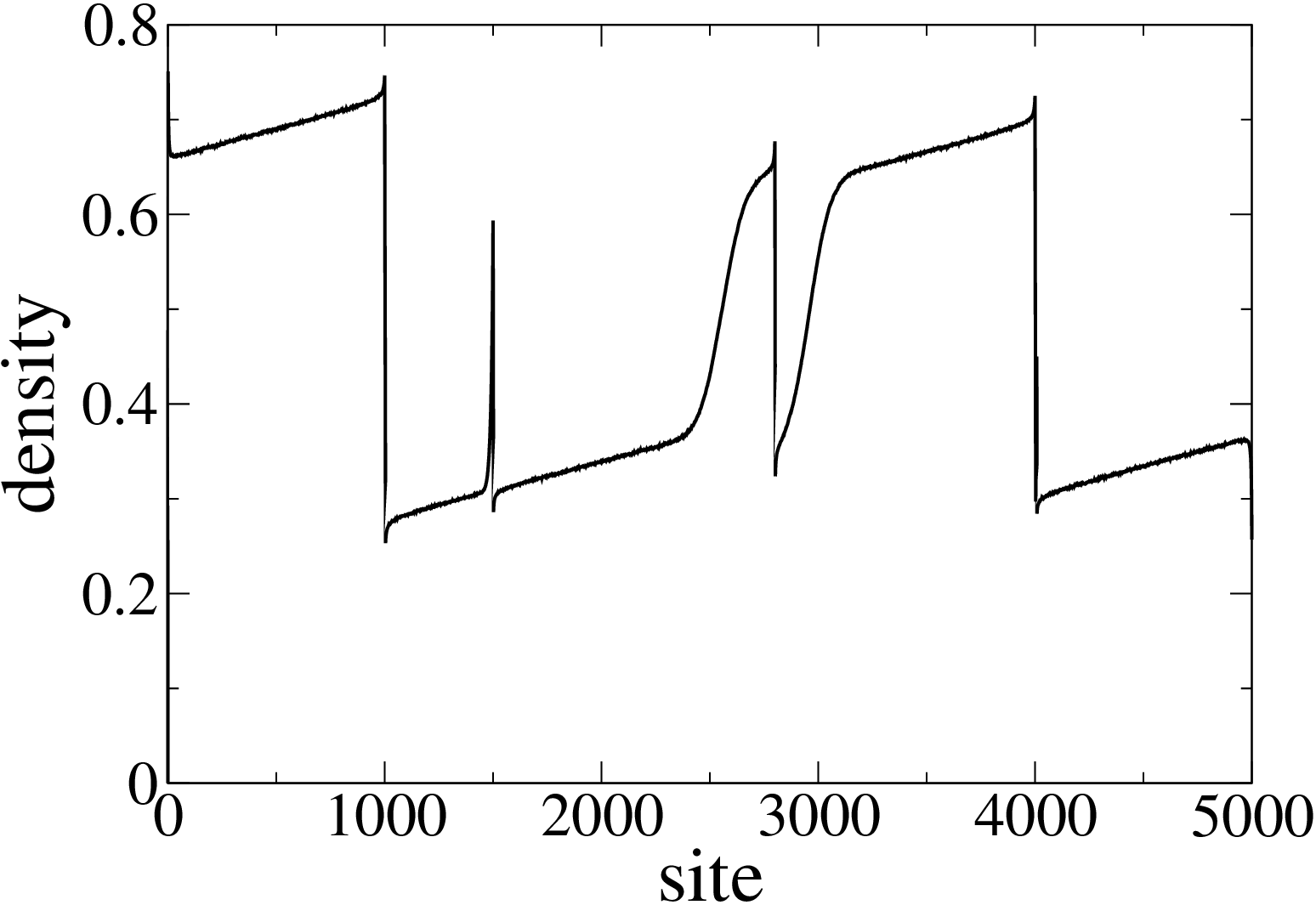}
\end{center}
\caption{\label{LIBA-check} (Colour online) Comparison of MC results and semi-analytical 
  results for the capacity field (= current profile for high boundary
  rates; here $\alpha=\beta=0.9$) by LIBA. Bottlenecks are at sites
  $x_i$ (first defect site) with size $l_1$:\\ $x_1=1000,\,\, l_1=4$ \\ 
  $x_2=1500,\,\, l_2=2$ \\ $x_3=2800,\,\, l_3=2$ \\ $x_4=4000,\,\,
  l_4=3$ \\ $x_5=4008,\,\, l_5=1$. \\ Further details are given in the
  text.}
\end{figure}

In Fig.~\ref{LIBA-check} we compare computer simulations of a system
with a few defects with results obtained by the minimal principle
(\ref{J(x)_glob}) in order to illustrate some features of the TASEP/LK
with defects. We chose high boundary rates, so that the 
resulting current profile is exactly the capacity field ${\mathcal C}(x)$. 
For the values $\Omega=\Omega_a=\Omega_d=0.2$
analytical results for the local current profiles in the continuum
limit are available. Following \cite{pff1,evans_pff} we used the reference
functions $\tilde J_L(x)=\Omega x-\Omega^2 x^2$ and $\tilde
J_H=-\Omega x+\Omega^2 x^2$ that obey the initial condition $\tilde
J_{L,H}(0)=0$ to reproduce the local solutions of (\ref{dJdx}). The
transport capacity was obtained in LIBA by results of a TASEP with a
single bottleneck. The first three bottlenecks are well separated by a
large distance. Here we see that LIBA works quite well and the current
profile is reproduced by the minimal principle quite accurately. We
also find that at the position of bottleneck 2, the actual current is
less than the transport capacity since the local solution of defect 1
is less than $J^*(x_2)$ \footnote{We observe a tiny spike at
the position of bottleneck 2, which is due to the influence of the
density peak on the slope of the current profile at this point,
though this effect should vanish in the continuum limit.}.
For bottleneck 4 there are deviations to LIBA since bottleneck 5 which
is quite close to bottleneck 4 (distance $=$ 6 sites) perturbs the
transport capacity by further decreasing it. Nonetheless also in this
region the minimal principle works if one takes the real transport
capacity \footnote{The value of the perturbed transport capacity at
  $x_4$ can actually be obtained by simulating a TASEP with a
  bottleneck of length 3 and a single defect at a distance of 6
  sites.} instead of LIBA.

\subsection{Local current profiles in the 
  disordered TASEP/LK}
\label{sec_loc_profs}

We now want to quantify our results by finding the local solutions of
the differential equation (\ref{dJdx}) and the continuity equation
(\ref{continuity_equ}), respectively. For a numerical evaluation of these
equations we need the CDR $J(\rho)$ and its inverse $\rho_{L,H}(J)$.

If there are only few defects in the system we have seen that the CDR
is the same as in the homogeneous system, as long as the current
is below the maximum current $J^*$, since the increase of the average
density is negligible. Thus in the TASEP/LK with defects we can use
the same CDR as in the homogeneous system: $J(\rho)=\rho(1-\rho)$.
Therefore the local solutions are the same as the ones of the
homogeneous systems.

The situation is different for a finite fraction of defects in the
system.  Then the average density is strongly influenced by the dense
distribution of defect peaks which leads to an altered current-density
relation even in the non-plateau region \cite{barma3}.  We will give
an approximation to calculate the current-density relation for small,
but finite, defect density $\phi\ll 1$ if it is not too close to the
maximum current. For that purpose we virtually divide the system into
homogeneous subsystems with fast hopping rate $p$, while the slow
hopping bonds connect these subsystems \footnote{This division into
  subdivision is motivated by the \emph{interacting subsystems
    approximation} (ISA) \cite{GreulichS08}}.  In first instance we
neglect correlations on the defect bonds.  The subsystems 
have an average size ${1}/{\phi}$. In this point 
of view, the peaks at the defects are the boundary layers of the 
homogeneous subsystems. Without losing generality, we can assume the 
system to be in the low density phase and observe the local solution 
of the right boundary where peaks are concave. 
This can be transfered to high density solutions by
particle-hole symmetry operation. Since $\omega_a,\omega_d\sim 1/L$ we
can neglect them for large systems when looking at a single subsystem,
thus we can treat them as homogeneous TASEPs.  In a large homogeneous
TASEP in the low density phase, the density is given by
$\rho_0=1/2-\sqrt{1/4-J}$ in the bulk far from the boundary. We can
write the \emph{mass} $m:=\sum_{i=1}^L \rho_i$ of the system as
$m=L\rho_0+m_p$ with $m_p$ being the mass of the boundary layer. $m_p$
thus corresponds to the mass of a peak in the inhomogeneous system.

\begin{figure}[ht]
\begin{center}
\vspace{0.5cm}
\includegraphics[height=5cm]{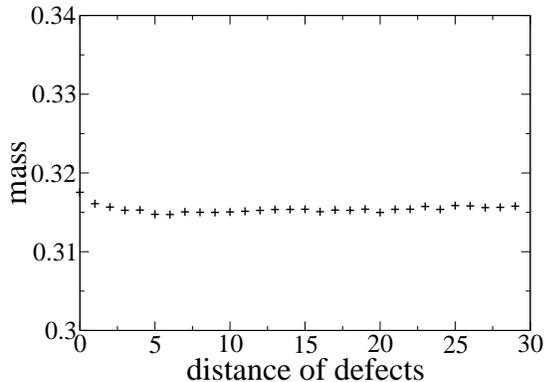}

\end{center}
\caption{\label{mass(defdist)}Mass $m$ of two density-peaks 
  $m=\sum_{i=1}^L (\rho_i-\alpha)$ in the low density phase of the
  TASEP with two defects in dependence on the distance of the defects.
  One observes that the dependence is rather weak.}
\end{figure}

We approximate that the mass of the peaks does not depend on distance
of adjacent defects.
Then we can write the average density as
\begin{equation}
\label{rho_(rho_0)}
\rho(x)=\rho_0(J(x))+\phi \, m_p(J(x)),
\end{equation}
since $\phi$ is the fraction of defect sites. Surprisingly this rather
uncontrolled approximation is supported by
Fig.~\ref{mass(defdist)} where we plotted the mass in a system with
two defects in dependence on the distance of latter ones. 

In this approximation, the mass of the peaks can be calculated
analytically, since due to the independence of distance we can take it
as the mass of the boundary layer in a large homogeneous TASEP,
where exact results are available for given current $J$ \cite{derrida1}. 
The density at a site $L-n$ is given by 
\begin{equation}
\label{peakdens}
\langle \tau_{L-n} \rangle=J S_n\left(J \right)+J^{n+1} 
  R_n\left(1/(1-\rho)\right)
\end{equation}
with 
\begin{eqnarray}
\label{S+R}
S_n(x) &=&\frac{1-\sqrt{1-4x}}{2x}-\sum_{j=n}^\infty \frac{(2j)!}{(j+1)!j!}, \\
R_n(x) &=& \sum_{j=2}^{n+1} \frac{(j-1)(2n-j)!}{n!(n+1-j)!}x^j\,.
\label{S+R2}
\end{eqnarray}
Thus the peak mass is
\begin{equation}
\label{peakmass}
m_p=\sum_n \left[\langle  \tau_{L-n}\rangle-\alpha(1-\alpha)\right]\,, 
\end{equation}
while the sum is truncated once the terms are small enough.

\begin{figure}[ht]
\begin{center}
\vspace{0.5cm}
\includegraphics[height=5cm]{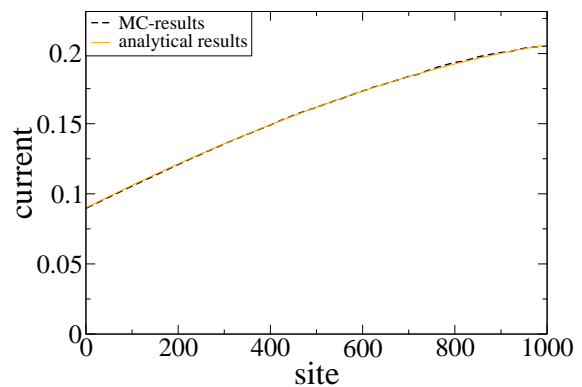}
\end{center}
\caption{\label{currprof_check} (Colour online) Comparison between simulation and 
  analytical results for the current profile in a disordered system
  with $\phi=0.2,\, \omega_a=0.2,\, \omega_d=0.1$ for entry rate
  $\alpha=0.1$ and exit rate $\beta=0.9$. Since the current is less
  than the transport capacity throughout the system the profile
  corresponds to the \emph{local} current profile of the boundary
  condition $\rho(0)=\alpha$. We observe excellent agreement between
  numerical and analytical results. This agreement holds for low
  current. Deviations occur only if the current comes close to the
  transport capacity.}
\end{figure}

Eqs.~(\ref{rho_(rho_0)})-(\ref{S+R2}) can be used to calculate the
current $J$ for a given density $\rho$ in the low density phase (and
in the high density phase by particle-hole symmetry) and vice versa:
\begin{equation}
J(\rho)=(\rho-\phi\, m_p)(1-\rho+\phi\, m_p)\,.
\end{equation}
This relation can be used to obtain a local solution of the
differential equation (\ref{dJdx}) for a given initial condition $J_i$
by iteration. In Fig.~\ref{currprof_check} we compared profiles
obtained by this procedure with results from computer simulations. One
observes an excellent agreement which holds if the current is not close to
the transport capacity. Together with the minimal current principle
(\ref{J(x)_glob}) the global current profile can be obtained.

The corresponding density profile can be obtained by inverting the CDR 
with respect to its two branches. Regions with a high density solution of the
current profile correspond to a high density domain with the density
$\rho_H(J(x))$ obtained by the inverted current density relation.
Analogous to that low density domains exist in regions of low density
solutions.

\subsection{Phase diagram of disordered systems}
\label{PD_sec}

We now want to investigate the phase diagram of inhomogeneous driven
lattice gases.

If one of the local boundary solutions $J_\alpha(x)$ or $J_\beta(x)$
is the minimum of all local solutions in the whole system, we have a
{\em low density phase} (L) in the former case and a {\em high density
  phase} (H) in latter one and there are no shocks in the system.
These phases have the same macroscopic properties like in the
corresponding homogeneous system.

If there are intersecting points of local solutions they manifest
themselves as shocks in the density profile, separating high and low
density regions (phase separation) corresponding to the realized high
and low density solutions of the current profile. Phase separation can
also be observed in homogeneous systems with Langmuir kinetics like
the TASEP/LK and the NOSC-model \cite{pff1,kif1a1,kif1a2}.
Here the local solutions of the boundaries $J_\alpha$ and $J_\beta$
can intersect leading to a single stationary shock in the density
profiles, separating a low density domain left of it and a high
density region right of it. This is called the {\em shock phase} (S)
\cite{pff1} which is preserved as long as the minimum local profiles
are the boundary current profiles. However, this kind of phase
separation differs from the phase separation induced by defects. While
in the S-phase the bulk behaviour is still determined by the boundary
conditions, phase separation due to the finite transport capacity of
defects is accompanied by a region where the current is ``screened''
by the defect(s) and is independent of the boundary condition, i.e.\ 
$\frac{\partial J(x)}{\partial \alpha}=0$ for all $x$ inside this
region. If the phase separation is due to the screening by defects we
rather refer to a \emph{defect-induced phase separated phase} (DPS).
If both boundary profiles $J_\alpha(x)$ and $J_\beta(x)$ are larger
than $\mathcal{C}(x)$ in the whole system, the complete system is
screened. The current profile is completely determined by the
defect distribution and identical to the capacity field $\mathcal{C}(x)$.
As argued in Sec.~\ref{model_sec} we call this fully screened phase 
\emph{Meissner phase (M)}.

Another possible scenario is that the current near the boundaries is only limited by the maximum
current of the bulk, i.e. ${\mathcal C}=J^{\rm max}_{\rm hom}$ and we have a \emph{maximum current
  phase} with long ranging boundary layers like in the homogeneous
TASEP. However in disordered systems with randomly disordered defects,
distances of defects are microscopic and the probability that 
${\mathcal C}=J^{max}_{\rm hom}$ vanishes in the continuum limit.  

We can characterize the
phases by two quantities:

\begin{enumerate}
\item The total length $\lambda_H$ of high density regions. This is
  the sum of individual high density regions and corresponds to the
  total jam length in traffic models \cite{traffic1}.

\item The screening length $\xi$ \footnote{This terminology is
    inspired by the screening length in \cite{pff_1def}. Nonetheless,
    the reader should be alert that in that work the meaning of $\xi$
    is different, corresponding to a \emph{maximum} screening length
    in our terminology}, which is the size of the area where the
  current profile does not depend on the boundary conditions. This is
  exactly the region where the boundary independent capacity field
  $\mathcal{C}(x)<J_{\alpha,\beta}(x)$ and the local boundary profiles
  are not the physically realized ones.
\end{enumerate}
In table \ref{phase table} the behaviour of these quantities in the
different phases is displayed. Indeed this can be used to {\em define} the
phases. For $\xi=0$ defects do not influence the current profile and
the system is in one of the ``pure'' phases, L,H or S, determined by the
boundary conditions. If $0<\xi<1$ there is phase separation and a part
of the system does not depend on the boundary conditions, the system
is in the DPS-phase. For $\xi=1$ the complete system is screened and
the current profile is solely determined by the defect
distribution and the system is in the M-phase. The ``pure'' phases
L,H,S can be characterized by $\xi=0$ and the vanishing of high
density regions (L, $\lambda=0$), coexistence of high and low density
regions (S, $0<\lambda<1$), and a global high density region (H,
$\lambda=1$). 

The transition from L or H to DPS is marked by a
discontinuity in $\xi$, but it is continuous in $\lambda$. Indeed due
to the discrete distribution of defects, $\xi$ itself is
discontinuous throughout the DPS-phase while $\lambda$ is not. In the
M-phase both $\xi$ and $\lambda$ are constant, while $\xi=1$ and
$\lambda$ takes a finite value $\lambda_M$ that is determined by the
fraction of high density regions in the capacity field
$\mathcal{C}(x)$ which depends on the individual defect distribution.

\begin{figure}[ht]
\begin{center}
\vspace{0.5cm}
\includegraphics[height=5cm]{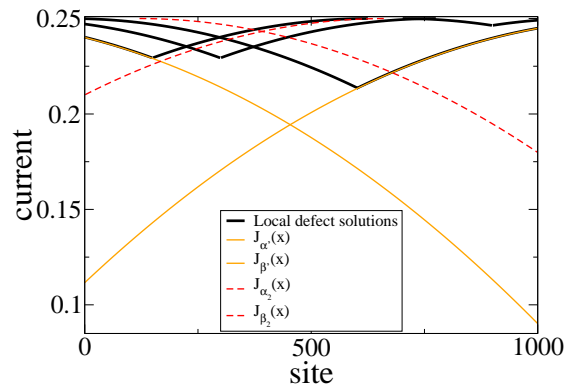}
\end{center}
\caption{\label{illust} 
  (Colour online) Illustration of some current profiles, including critical profiles.
  We see that the critical rates are related by the critical current
  profiles: $\alpha^*=\rho_L(J_{\beta'}(1))$, while
  $\rho_L(\rho)$ is the inverted (low density) CDR and
  $J_{\beta'}(1)$ is the local right boundary solution for
  $\beta=\beta'$. An analog relation is valid for $\beta^*$. The
    bold lines are the local current profiles consistent with the
    initial conditions imposed by the defects, whose minimal envelope
    is the capacity field. The thin lines are the critical boundary
    profiles and the dashed line corresponds to phase separated
    boundary current profiles.  }
\end{figure}

\begin{table}
\begin{center}
\begin{tabular}{|c|c|c|c|c|c|}
  \hline & L & H & S & DPS & M \\ \hline $\lambda$ & 0 & 1 &
  $\begin{array}{c} 0<\lambda<1 \\ \mbox{continuous}
\end{array}$ &
$\begin{array}{c}
0<\lambda<1 \\ 
\mbox{continuous}
\end{array}$ 
& $\lambda_M$ \\
\hline
$\xi$ & 0 & 0 & 0 &
$\begin{array}{c}
0<\xi<1 \\ 
\mbox{discontinuous}
\end{array}$ & 1 \\
\hline
\end{tabular}
\end{center}
\caption{\label{phase table} Values and properties of the characteristic 
  order parameters $\xi$ and $\lambda$ in the different phases. These
  properties can be used to define phases.}
\end{table}

We see that at most phase boundaries both quantities $\xi$ and
$\lambda$ are non-analytic. At the transition from S to DPS though
$\lambda$ is analytic, thus it cannot be characterized by $\lambda$.
Hence for theoretical investigations it appears to be more convenient
to use $\xi$ to discriminate defect- and non-defect phases. In
simulations it is easier to detect phase separation (see next
section) and use the non-analytic behaviour of $\lambda$ to obtain
critical points. Due to the analytic behaviour between S- and
DPS-phase, however, this approach is only applicable at L-DPS and
H-DPS-transitions. The S-DPS transition has to be obtained by
theoretical considerations.

In particle-conserving systems with defects the DPS- and S-phases
vanish since no stationary shocks are possible. Here both $\xi$ and
$\lambda$ are discontinuous at the transition to the M-phase. However,
in these systems the Meissner phase usually is also called ``phase
separated phase'' \cite{GreulichS08} since no distinction between
\emph{several} phases with phase separation has to be made.

\begin{figure}[ht]
\begin{center}
\vspace{0.5cm}
\includegraphics[width=0.99\columnwidth]{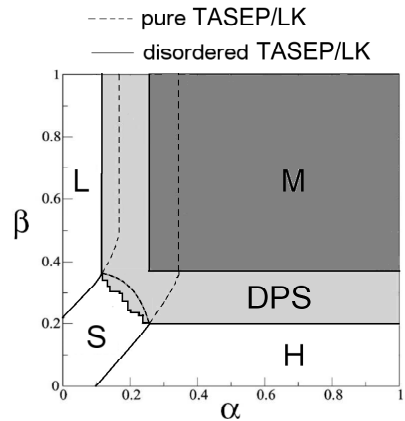}
\end{center}
\caption{\label{PD2} Phase diagram of the disordered TASEP/LK for 
  $\Omega_a>\Omega_d$. The critical rates depend on each other as
  $\alpha^*=\rho_L(J_\beta'(1))$, which is argued in the text. The
  transition line between S- and DPS-phase is not smooth in the weak
  continuum limit due to the unsmooth structure of the capacity field
  (bold line). In the strong continuum limit the DPS phase is concave
  (bold dashed line). The topology of other disordered driven lattice
  gases is expected to be the same.}
\end{figure}

A sketch of the $\alpha-\beta$ phase diagram of a disordered driven lattice gas with LK is
displayed in Fig.~\ref{PD2}. Attachment and detachment rates are
fixed, while here $\omega_d>\omega_a$. L-,H- and even S-phase might vanish for
large $\Omega_{a,d}$ if $J_{\alpha=0}(L)>J^*(x_b)$ at some point $x_d$
for any boundary rate $\alpha$ or $\beta$ so that phase separation
with screening already occurs for vanishing boundary density. The
dashed lines mark the phases of the homogeneous system.  These pure
phases are overlayed by the DPS- and M- phase which are characterized
by the critical boundary rates $\alpha', \beta'$ and
$\alpha^*,\,\beta^*$. $\alpha'$ and $\beta'$ mark the minimal boundary
rates at which the respective local boundary profile intersects the
capacity field, i.e $J_{\alpha,\beta}>\mathcal{C}(x)$ for at least one
point $x$, while at the rates $\alpha^*,\,\beta^*$,
$J_{\alpha,\beta}>\mathcal{C}$ everywhere, so that local boundary 
profiles cannot propagate into the bulk. In Fig.~\ref{illust} we sketched some
critical current profiles to illustrate the critical parameters. In
parameter regions where $J_\alpha$ and $J_\beta$ do not intersect,
$\alpha'$ and $\beta'$ do not depend on each other as well as
$\alpha^*$ and $\beta^*$, hence the phase diagram has a simple
structure with phase boundaries parallel to the parameter axes.
Though, as we can see in Fig.~\ref{illust}, $\alpha'$ and $\beta^*$ do
depend on each other since $J(\beta^*)=J_{\alpha'}(1)$. The same
relation is valid for $\beta'$ and $\alpha^*$.  Inside the region of
intersecting boundary profiles (the shock phase of the homogeneous
system), the structure is nontrivial. The phase transition betweeen S-
and DPS-phase depends explicitely on the variation of the intersection
points of boundary profiles and minimal defect profiles.  Explicitely
it is given by the condition that a triple points $x_t$ with
$J_\alpha(x_t)=J_\beta(x_t)=\mathcal{C}(x)$ exist. One special case
for which this condition can be solved exactly is the disordered TASEP/LK for
$\Omega_a=\Omega_d$ in the strong continuum limit, where terms of
${\mathcal O}(1/\ln L)$ are neglected and the defect density $\phi$
scaling to zero as $\phi\sim 1/\ln L$. In this case the capacity field
${\mathcal C}$ is constant and the transition line is just a diagonal
straight line. The phase diagram in the strong continuum limit
is derived in the Appendix and displayed in Fig.~\ref{PD1}. Though 
this limit is not quite physical it can be used as a reference point to 
argue that for finite defect densities the S-phase is convex 
(see also the Appendix).

If we go away from the strong continuum limit, $\mathcal{C}(x)$ 
is not a constant. The structure of $\mathcal{C}$ is not smooth as was argued
in Sec.~\ref{global_sec}, so is the transition line. In
Fig.~\ref{PD2} we displayed a rather
generic sketch of a phase diagram that incorporates these arguments.
Phase diagrams of other driven lattice gases with the properties noted
in the introduction will have the same topology.
 
\begin{figure}[ht]
\begin{center}
\vspace{0.5cm}
\includegraphics[width=0.8\columnwidth]{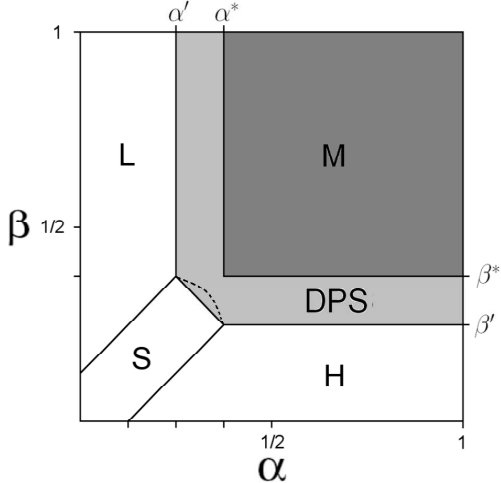}
\end{center}
\caption{\label{PD1} Phase diagram of the disordered TASEP/LK with 
  $\Omega_a=\Omega_d=:\Omega$ in the strong continuum limit. The bold
  line at the S-DPS-boundary is valid for $\phi$ scaling as $1/\ln L$ and
  dashed line (sketched) is valid for finite defect density. The
  critical rates are given by $\alpha^*=\beta^*=(1-\sqrt{1-q})/2$,
  $\alpha'=\alpha^*-\Omega$, same for $\beta$, with $\Omega=0.1,\,\,
  q=15/16$. The phase boundaries of the S-phase are of second order.
  For $\Omega>1/2$, L- and H-phase vanish.}
\end{figure}


\section{Expectation values for phase transitions}
\label{sec-expval}

Like in particle conserving systems, the properties of disordered
driven lattice gases with Langmuir kinetics depend strongly on
microscopic details of the defect sample. Since we are interested in
macroscopic properties that do not depend on microscopic defect
distributions, we concentrate on probabilistic quantities of ensembles
of systems. One quantity of interest is the expected fraction of
systems that exhibit phase separation in an ensemble of systems with
identical system parameters and defect density. In this section we
derive a procedure to calculate this quantity based on analytical
results obtained by the principles from the last section.

In order to compare these results with Monte Carlo simulations we
introduce virtual particles similiar to second class
particles \cite{scp1} that indicate if phase separation
occurs in the simulated system. These particles do not change the
dynamics of the system. The predicted probability for phase separation
is then compared with the relative frequency of phase separation in a
set of simulations.

\subsection{Automated detection of phase separation}
\label{V-particle_sec}

We introduce so-called {\em virtual particles} (V-particles) to identify and
distinguish high and low density regions. These particles do not
follow the exclusion constraint, instead they can occupy all sites
even if these are occupied by particles. The dynamics of the
V-particles is the following: At the beginning, a V-particles is put
on each defect site. After each lattice update the
V-particles are updated sequentially beginning at the left. Each
V-particle hops to the right if there is a particle on its site, while
it hops to its left adjacent site if it is residing on an empty site.
The V-particle cannot hop over slow bonds, thus if it is on a defect
site, it cannot hop to the right, while if it is on a site right of a
defect site, it cannot hop to the left. Hence, at any time, there is
exactly one V-particle between each pair of contiguous defect sites.
If the average density between two
defects is larger than $1/2$, the V-particle tends to move to the
right, while for $\rho<1/2$ it tends to move to the left. Thus, we can
identify a high density region by a V-particle that is, on average,
closer to the right defect.  By computing the average distance of a
V-particle to the defect right of it we can identify if there is a
high density region in its vicinity.

Using this procedure we can run a large number of simulations and
automatically identify whether high and low density regions coexist.
In this way the relative frequency of phase separated systems and an 
estimate for the probability of phase separation can be determined.

\subsection{Analytical approach for phase separation probability}

We use the results from the last sections in order to derive a
analytical approach that allows the determination of the probability
that for a given defect density $\phi$ phase separation occurs. Again
we consider ensembles of systems instead of a fixed configuration
of defects.

The condition that no phase separation occurs is 
\begin{equation}
\label{cond_noPS_2}
J_\alpha(x)<J^*(x_b) \mbox{\ \ and\ \ } J_\beta(x)<J^*(x_b) \quad
\mbox{for all $x_b$}.
\end{equation}

The fact that only low density solutions can intersect high density
solutions also implies that an increases of $\alpha$ leads to a left
shift of phase boundaries (in the phase separated phase) to the left
while a increase of $\beta$ moves the phase boundaries to the right.
This can be seen in Fig.~\ref{phase_sep}.

Following the LIBA we assume that the transport capacity at a position
$x$ approximately depends only on the length of the bottleneck at this
point, thus $J^*(x)\approx J^*(l(x))$.  
In a system with binary disorder there are on
average $L(1-\phi)$ bottlenecks and the probability that one specific
bottleneck has length $l$ is $P(l)=(1-\phi)\phi^l$ \cite{GreulichS08a}.

The relation between bottleneck length and
transport capacity $J^*(l)$ as well as its inverse relation
$l(J^*)$  can be obtained by analytical considerations or numerical 
computations in single bottleneck systems. The probability that the current 
is below the transport
capacity at a given position $x$ is then 
\begin{equation}
  P(J<J^*)=P(l<l(J))=\sum_{l'=0}^{\lfloor l(J) \rfloor}
  P(l)=1-\phi^{\lfloor l(J(x)) \rfloor} \,\,\, .
\end{equation}

\begin{figure}[ht]
\begin{center}
\vspace{0.5cm}
\includegraphics[height=5cm]{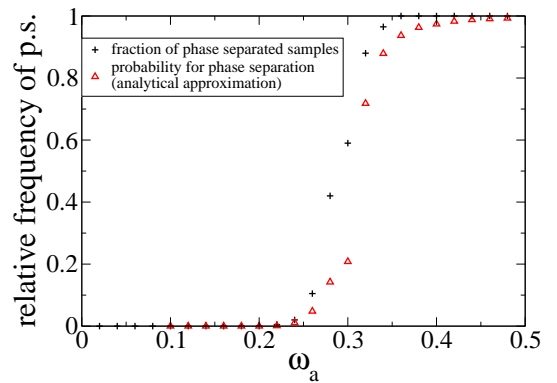}
\end{center}
\caption{\label{freq(PS)_wa} (Colour online) Fraction of samples that
  exhibit phase separation in dependence of the attachment rate
  $\omega_a$ for fixed $\alpha=0.1,\, \beta=0.9,\, \omega_d=0.3$. The
  system size is $L=1000$ and each data point is obtained by
  simulating 200 random defect samples with same system parameters.
  This is compared with analytical results obtained by
  (\ref{P(noPS)}).}.
\end{figure}

The probability $\mathcal P$ that no phase separation occurs is equal
to the probability that the current is below the transport capacity
everywhere in the system:
\begin{equation}
\label{P(noPS)}
{\mathcal P}=\prod_{i=1}^{\langle N_{bn}\rangle}
P(J(i)<J^*(l(i))=\prod_{i=1}^L (1-\phi^{\lfloor l(J(i))\rfloor}) \,\, .
\end{equation}
Here $N_{bn}$ is the number of bottlenecks from left to right, so
$J(i)$ is the current at bottleneck $i$ counted from the left. Since
on average there are $\langle N_{bn}\rangle=L(1-\phi)$ bottlenecks, we
can determine $J(i)$ recursively by rescaling eq.~(\ref{continuity_equ})
by the factor $1/(1-\phi)$ to obtain
\begin{eqnarray}
  J(i+1) &=& J(i)+\omega_a(1-\phi)(1-\rho(i))-\omega_d(1-\phi)\rho(i) \,, 
          \nonumber\\ 
\rho(i) &=& \rho_0(J(i))+\phi m_p \,\,.
\end{eqnarray}
This way the probability for phase separation, which explicitely
depends on the system size can be computed iteratively by
(\ref{P(noPS)}), while analytical results for $J(l)$ in the TASEP with
a single bottleneck are available \cite{GreulichS08}. In comparison to
Monte Carlo simulations, this computation can be made with little
effort. In Fig.~\ref{freq(PS)_wa} we simulated ensembles of random
defect samples for different parameter values. The fraction of samples
exhibiting phase separation is determined by the method from
subsection~\ref{V-particle_sec} and compared with results obtained by
(\ref{P(noPS)}). One observes a region with a quite steep increase of
the probability. The analytical results fit the simulation results
quite nicely, although there is a small shift to larger values of
$\omega_a$.
 

\section{Summary and conclusions}
\label{sec-summary}

In this paper we have investigated the interplay between Langmuir kinetics
(particle creation and annihilation in the bulk) and disorder, realized
through randomly distributed hopping rates, in driven lattice
gases connected to boundary reservoirs.
Although both features provide a mechanism for phase separation (shock
formation), the underlying mechanisms and dynamics is different and 
might lead to a form of competition. 

Based on Monte Carlo simulations of the disordered TASEP/LK, a TASEP
with Langmuir kinetics and site-disordered hopping rates, the main
properties of such systems have been identified. Like in the
disordered TASEP we observe narrow peaks in the vicinity of defect 
sites. Their width however vanishes in the continuum limit. 
For larger values of the boundary rates we observe 
\emph{defect-induced phase separation}, where multiple macroscopic high 
and low density regions with a \emph{multitude} of shocks occur. 

These findings can be understood in terms of an extremal principle. 
In contrast to the principle originally proposed for homogeneous systems
\cite{popkov2,kolomeisky_shockdyn} it is a \emph{local} principle for
the \emph{current profile}. This is a direct consequence of the
interplay between Langmuir kinetics, which induces a site-dependence
of the stationary current, and the randomly distributed
inhomogeneities. In our approach we assumed that defects locally 
induce a reduced transport capacity imposing an upper bound to the
current. For weakly interacting systems this quantity approximately
depends only on the local distribution of defects, especially on the size
of the bottleneck. In this approximation (LIBA) we can obtain the
transport capacity by refering to single bottleneck systems. The
transport capacity provides additional initial conditions to the
differential equation (\ref{dJdx}) that gives the slope of the local
current profile in the continuum limit, each of them representing a
individual local solution.  Shock dynamics impose additional
conditions on the physical current profile. Hence, out of the
multitude of solutions only the profile that locally minimizes all
solutions is physically realized.  
The full current profile can be obtained by superposing the solutions of 
all single bottlenecks which are described in terms of the
same ``Green's function'' $\tilde J_{L,H}(x)$ defined
in Sec.~\ref{global_sec}.

While in systems with only few defects local current profiles are
almost identical to those of the homogeneous systems, they
significantly differ in large systems with a finite fraction of defect
sites. In the case of the disordered TASEP/LK local density profiles
can be accurately reproduced by identifying the density peaks with
boundary layers of small virtual subsystems where exact results
are available. 

The minimal principle can be used to predict some features of the
phase diagram. As was already observed for single defects in
\cite{pff_1def}, defects can generate screened regions where the
influence of boundary conditions vanishes. 
We can distinguish the original non-screened phases
which are also present in homogeneous systems, a partially screened
phase exhibiting phase separation and a fully screened phase where the
influence of the boundary conditions vanishes completely. For the
strong continuum limit where terms of ${\mathcal O}(1/\ln L)$ do not
contribute, the minimal principle even allows the determination of the
exact phase diagram, while in for the weak continuum limit at least
most qualitative aspects of the phase diagram remain accessible.

The LIBA and the minimal principle can also be applied together with a
statistical approach to obtain an approximation for the probability
that a randomly produced disorder sample exhibits phase separation.

Although the results have been derived and tested on the TASEP/LK
we believe that are generic for a large class of driven lattice
gases, at least if they are ergodic with short-ranged interactions
and a single maximum in the current-density relation.  
In more general processes the lateral current  $s(\rho)$ takes the
role of attachment/detachment processes.


\section*{Acknowledgment}
\noindent

We thank Ludger Santen for useful discussions.


\appendix
\section*{APPENDIX: Phase transition lines in the strong continuum limit}
\label{appendixA}

Usually it is quite difficult to determine the transition line between
S- and DPS-phase. One special case where it is possible to solve that
problem exactly is in the strong continuum limit in the disordered
TASEP/LK for $\Omega_a=\Omega_d=:\Omega$. In addition the number of
defects is infinite, while the defect density is scaled to zero as
$\phi=\mathcal{O}(1/\ln L)$. The average length of the longest
bottleneck in a system of size $L$ scales as $\ln L/\ln \phi$
\cite{DTASEP_krug,GreulichS08a}, so in the strong continuum limit
there has to be an infinitely large bottleneck with a local
transport capacity $J^*_M=q/4$.  Moreover, we can say that this is the
case for any small interval of length $\varepsilon$ if $\varepsilon$
is scaling slower than $\sqrt{1/L}$, corresponding to
$L\sqrt{1/L}=\sqrt{L}$ sites. The global capacity field therefore
simply is the constant function $\mathcal{C}(x)=q/4$. Since the defect
density vanishes, the CDR is the same as in the homogeneous system as
was shown in the last sections numerically and analytically. The local
boundary current and density profiles will therefore be the same as in
the homogeneous system. Now the problem we have to solve is equivalent
to finding the transition from S- to LMH-phase in the homogeneous
TASEP/LK if the homogeneous maximum current $J^*=1/4$ is exchanged by
$q/4$ \cite{pff1,evans_pff}. In these works, the transition line was
determined to be
$\tilde\beta^*(\tilde\alpha)=\rho_L(J^*)-\Omega-\tilde\alpha$.
Inserting $J^*=q/4$, we obtain for the transition line
\begin{equation}
\label{pt_line}
\tilde\beta^*(\tilde\alpha)=1/2-\sqrt{\frac{1-q}{4}}-\Omega-\tilde\alpha 
\end{equation}
which is just a shift of the phase transition line to the right by the
term $\sqrt{(1-q)/4}$. The properties of the phases of course are
different to the ones in the homogeneous system as we have argued before
(especially the absence of long ranged boundary layers). The phase
diagram is displayed in Fig.~\ref{PD1}. We have to point out that in
this limit, the transition is of second order, since $\xi$ is
continuous.

Nonetheless the vanishing of $\phi$ in the continuum limit is not
quite physical, so we try to obtain at least qualitative results for the S-DPS transition line for finite $\phi$. In sec. \ref{sec_MC_finphi} and \ref{sec_loc_profs} we have seen that a small but finite defect density $\phi>0$ leads to a flattening of the local density profiles due to a broadening of the density peaks, so that their slopes $\frac{\partial \rho_{L,H}}{\partial x}$, which are positive for $\Omega_a=\Omega_d, \,\, \alpha<1/2,\,\beta<1/2$, are decreasing for higher current $J$.

Assume the system is on the transition line between S and DPS, i.e. a triple point 
$x_t$ with $J_\alpha(x_t)=J_\beta(x_t)=q/4$ exists. A shift of both $\rho_\alpha$ 
and $\rho_\beta$ by an infinitisemal amount $dx$ also 
shifts the triple point though it persists. In parameter space, this corresponds to a 
movement along the transition line, while the boundary values are changed by 
\begin{eqnarray}
d\alpha &=& \left.\frac{\partial \rho_\alpha}{\partial x}\right|_{x=0} dx \,\,\,\mbox{ and }\,\,\, d\beta=-\left.\frac{\partial \rho_\beta}{\partial x}\right|_{x=1} dx \\
&\Rightarrow& \frac{d\beta}{d\alpha}=-\frac{\left.\frac{\partial \rho_\beta}{\partial x}\right|_{x=1}}{\left.\frac{\partial \rho_\alpha}{\partial x}\right|_{x=0}}
\end{eqnarray}
using the relations $\alpha=\rho_\alpha(0)$ and $\beta=1-\rho_\beta(1)$. Since the boundary current $J_{\alpha,\beta}$ is monotonously increasing with $\alpha$ and $\beta$ for $\alpha,\beta<1/2$, the flattening of the density profiles leads to:
\begin{eqnarray}
\mbox{For } \beta>\alpha :  \left.\frac{\partial \rho_\beta}{\partial x}\right|_{x=1}<\left.\frac{\partial \rho_\alpha}{\partial x}\right|_{x=0} & \Rightarrow &\,\, \frac{d\beta}{d\alpha}>-1 \\ 
\mbox{For } \alpha>\beta : \left.\frac{\partial \rho_\beta}{\partial x}\right|_{x=1}>\left.\frac{\partial \rho_\alpha}{\partial x}\right|_{x=0} & \Rightarrow &\,\, \frac{d\beta}{d\alpha}<-1
\end{eqnarray}
along the transition line. This corresponds to a concave distortion of the DPS-phase as displayed in Fig. \ref{PD1}.



\begin{thebibliography}{40}
\expandafter\ifx\csname natexlab\endcsname\relax\def\natexlab#1{#1}\fi
\expandafter\ifx\csname bibnamefont\endcsname\relax
  \def\bibnamefont#1{#1}\fi
\expandafter\ifx\csname bibfnamefont\endcsname\relax
  \def\bibfnamefont#1{#1}\fi
\expandafter\ifx\csname citenamefont\endcsname\relax
  \def\citenamefont#1{#1}\fi
\expandafter\ifx\csname url\endcsname\relax
  \def\url#1{\texttt{#1}}\fi
\expandafter\ifx\csname urlprefix\endcsname\relax\def\urlprefix{URL }\fi
\providecommand{\bibinfo}[2]{#2}
\providecommand{\eprint}[2][]{\url{#2}}

\bibitem[{\citenamefont{Tripathy and Barma}(1997{\natexlab{a}})}]{barma1}
\bibinfo{author}{\bibfnamefont{G.}~\bibnamefont{Tripathy}} \bibnamefont{and}
  \bibinfo{author}{\bibfnamefont{M.}~\bibnamefont{Barma}},
  \bibinfo{journal}{Phys. Rev. Lett.} \textbf{\bibinfo{volume}{78}},
  \bibinfo{pages}{3039} (\bibinfo{year}{1997}{\natexlab{a}}).

\bibitem[{\citenamefont{Tripathy and Barma}(1998{\natexlab{b}})}]{barma2}
\bibinfo{author}{\bibfnamefont{G.}~\bibnamefont{Tripathy}} \bibnamefont{and}
  \bibinfo{author}{\bibfnamefont{M.}~\bibnamefont{Barma}},
  \bibinfo{journal}{Phys. Rev. E} \textbf{\bibinfo{volume}{58}},
  \bibinfo{pages}{1911} (\bibinfo{year}{1998}{\natexlab{b}}).

\bibitem[{\citenamefont{Shaw et~al.}(2004)\citenamefont{Shaw, Kolomeisky, and
  Lee}}]{ShawKL2004}
\bibinfo{author}{\bibfnamefont{L.}~\bibnamefont{Shaw}},
  \bibinfo{author}{\bibfnamefont{A.}~\bibnamefont{Kolomeisky}},
  \bibnamefont{and} \bibinfo{author}{\bibfnamefont{K.}~\bibnamefont{Lee}},
  \bibinfo{journal}{J. Phys. A} \textbf{\bibinfo{volume}{37}},
  \bibinfo{pages}{2105} (\bibinfo{year}{2004}).

\bibitem[{\citenamefont{Chou and Lakatos}(2004{\natexlab{a}})}]{ChouL2004b}
\bibinfo{author}{\bibfnamefont{T.}~\bibnamefont{Chou}} \bibnamefont{and}
  \bibinfo{author}{\bibfnamefont{G.}~\bibnamefont{Lakatos}},
  \bibinfo{journal}{Phys. Rev. Lett.} \textbf{\bibinfo{volume}{93}},
  \bibinfo{pages}{198101} (\bibinfo{year}{2004}{\natexlab{a}}).

\bibitem[{\citenamefont{Lakatos et~al.}(2006)\citenamefont{Lakatos, O'Brien,
  and Chou}}]{LakatosBC.2006}
\bibinfo{author}{\bibfnamefont{G.}~\bibnamefont{Lakatos}},
  \bibinfo{author}{\bibfnamefont{J.}~\bibnamefont{O'Brien}}, \bibnamefont{and}
  \bibinfo{author}{\bibfnamefont{T.}~\bibnamefont{Chou}}, \bibinfo{journal}{J.
  Phys. A} \textbf{\bibinfo{volume}{39}}, \bibinfo{pages}{2253}
  (\bibinfo{year}{2006}).

\bibitem[{\citenamefont{Barma}(2006)}]{barma3}
\bibinfo{author}{\bibfnamefont{M.}~\bibnamefont{Barma}},
  \bibinfo{journal}{Physica A} \textbf{\bibinfo{volume}{372}},
  \bibinfo{pages}{22} (\bibinfo{year}{2006}).

\bibitem[{\citenamefont{Juh\'asz et~al.}(2006)\citenamefont{Juh\'asz, Santen, and
  Igl\'oi}}]{PASEP_dis_santen}
\bibinfo{author}{\bibfnamefont{R.}~\bibnamefont{Juh\'asz}},
  \bibinfo{author}{\bibfnamefont{L.}~\bibnamefont{Santen}}, \bibnamefont{and}
  \bibinfo{author}{\bibfnamefont{F.}~\bibnamefont{Igl\'oi}},
  \bibinfo{journal}{Phys. Rev. E} \textbf{\bibinfo{volume}{74}},
  \bibinfo{pages}{061101} (\bibinfo{year}{2006}).

\bibitem[{\citenamefont{Dong et~al.}(2007)\citenamefont{Dong, Schmittmann, and
  Zia}}]{prot_prod_zia2}
\bibinfo{author}{\bibfnamefont{J.}~\bibnamefont{Dong}},
  \bibinfo{author}{\bibfnamefont{B.}~\bibnamefont{Schmittmann}},
  \bibnamefont{and} \bibinfo{author}{\bibfnamefont{R.}~\bibnamefont{Zia}},
  \bibinfo{journal}{J. Stat. Phys.} \textbf{\bibinfo{volume}{128}},
  \bibinfo{pages}{21} (\bibinfo{year}{2007}).

\bibitem[{\citenamefont{Foulaadvand et~al.}(2007)\citenamefont{Foulaadvand,
  Chaaboki, and Saalehi}}]{DTASEP_num}
\bibinfo{author}{\bibfnamefont{M.~E.} \bibnamefont{Foulaadvand}},
  \bibinfo{author}{\bibfnamefont{S.}~\bibnamefont{Chaaboki}}, \bibnamefont{and}
  \bibinfo{author}{\bibfnamefont{M.}~\bibnamefont{Saalehi}},
  \bibinfo{journal}{Phys. Rev. E} \textbf{\bibinfo{volume}{75}},
  \bibinfo{pages}{011127} (\bibinfo{year}{2007}).

\bibitem[{\citenamefont{Greulich and
  Schadschneider}(2008{\natexlab{a}})}]{GreulichS08}
\bibinfo{author}{\bibfnamefont{P.}~\bibnamefont{Greulich}} \bibnamefont{and}
  \bibinfo{author}{\bibfnamefont{A.}~\bibnamefont{Schadschneider}},
  \bibinfo{journal}{Physica A} \textbf{\bibinfo{volume}{387}},
  \bibinfo{pages}{1972} (\bibinfo{year}{2008}{\natexlab{a}}).

\bibitem[{\citenamefont{Greulich and
  Schadschneider}(2008{\natexlab{b}})}]{GreulichS08a}
\bibinfo{author}{\bibfnamefont{P.}~\bibnamefont{Greulich}} \bibnamefont{and}
  \bibinfo{author}{\bibfnamefont{A.}~\bibnamefont{Schadschneider}},
  \bibinfo{journal}{J. Stat. Mech.} p. \bibinfo{pages}{P04009}
  (\bibinfo{year}{2008}{\natexlab{b}}).

\bibitem[{\citenamefont{Grzeschik et~al.}(2008)\citenamefont{Grzeschik, Harris,
  and Santen}}]{GrzeschikHS08}
\bibinfo{author}{\bibfnamefont{H.}~\bibnamefont{Grzeschik}},
  \bibinfo{author}{\bibfnamefont{R.}~\bibnamefont{Harris}}, \bibnamefont{and}
  \bibinfo{author}{\bibfnamefont{L.}~\bibnamefont{Santen}},
  \bibinfo{journal}{arXiv:0806.3845}  (\bibinfo{year}{2008}).

\bibitem[{\citenamefont{Krug}(2000)}]{DTASEP_krug}
\bibinfo{author}{\bibfnamefont{J.}~\bibnamefont{Krug}}, \bibinfo{journal}{Braz.
  Jrl. Phys.} \textbf{\bibinfo{volume}{30}}, \bibinfo{pages}{97}
  (\bibinfo{year}{2000}).

\bibitem[{\citenamefont{Enaud and Derrida}(2004)}]{enaud}
\bibinfo{author}{\bibfnamefont{C.}~\bibnamefont{Enaud}} \bibnamefont{and}
  \bibinfo{author}{\bibfnamefont{B.}~\bibnamefont{Derrida}},
  \bibinfo{journal}{Europhys. Lett.} \textbf{\bibinfo{volume}{66}},
  \bibinfo{pages}{83} (\bibinfo{year}{2004}).

\bibitem[{\citenamefont{Harris and Stinchcombe}(2004)}]{harris_MF}
\bibinfo{author}{\bibfnamefont{R.~J.} \bibnamefont{Harris}} \bibnamefont{and}
  \bibinfo{author}{\bibfnamefont{R.~B.} \bibnamefont{Stinchcombe}},
  \bibinfo{journal}{Phys. Rev. E} \textbf{\bibinfo{volume}{70}},
  \bibinfo{pages}{016108} (\bibinfo{year}{2004}).

\bibitem[{\citenamefont{Jain and Barma}(2003)}]{jain_DZRP}
\bibinfo{author}{\bibfnamefont{K.}~\bibnamefont{Jain}} \bibnamefont{and}
  \bibinfo{author}{\bibfnamefont{M.}~\bibnamefont{Barma}},
  \bibinfo{journal}{Phys. Rev. Lett.} \textbf{\bibinfo{volume}{91}},
  \bibinfo{pages}{135701} (\bibinfo{year}{2003}).

\bibitem[{\citenamefont{Stinchcombe}(2002)}]{disorder_stinche}
\bibinfo{author}{\bibfnamefont{R.}~\bibnamefont{Stinchcombe}},
  \bibinfo{journal}{J. Phys.: Condens. Matter} \textbf{\bibinfo{volume}{14}},
  \bibinfo{pages}{1473} (\bibinfo{year}{2002}).

\bibitem[{\citenamefont{Derrida et~al.}(1993)\citenamefont{Derrida, Evans,
  Hakim, and Pasquier}}]{derrida1}
\bibinfo{author}{\bibfnamefont{B.}~\bibnamefont{Derrida}},
  \bibinfo{author}{\bibfnamefont{M.~R.} \bibnamefont{Evans}},
  \bibinfo{author}{\bibfnamefont{V.}~\bibnamefont{Hakim}}, \bibnamefont{and}
  \bibinfo{author}{\bibfnamefont{V.}~\bibnamefont{Pasquier}},
  \bibinfo{journal}{J. Phys. A} \textbf{\bibinfo{volume}{26}},
  \bibinfo{pages}{1493} (\bibinfo{year}{1993}).

\bibitem[{\citenamefont{Sch\"utz and Domany}(1993)}]{schuetz_dom}
\bibinfo{author}{\bibfnamefont{G.}~\bibnamefont{Sch\"utz}} \bibnamefont{and}
  \bibinfo{author}{\bibfnamefont{E.}~\bibnamefont{Domany}},
  \bibinfo{journal}{J. Stat. Phys.} \textbf{\bibinfo{volume}{72}},
  \bibinfo{pages}{277} (\bibinfo{year}{1993}).

\bibitem[{\citenamefont{Blythe and Evans}(2007)}]{BlytheE07}
\bibinfo{author}{\bibfnamefont{R.}~\bibnamefont{Blythe}} \bibnamefont{and}
  \bibinfo{author}{\bibfnamefont{M.}~\bibnamefont{Evans}}, \bibinfo{journal}{J.
  Phys. A} \textbf{\bibinfo{volume}{40}}, \bibinfo{pages}{R333}
  (\bibinfo{year}{2007}).

\bibitem[{\citenamefont{Lodish et~al.}(2003)\citenamefont{Lodish, Berk,
  Matsudaira, Kaiser, Krieger, Scott, Zipursky, and Darnell}}]{CellBook}
\bibinfo{author}{\bibfnamefont{H.}~\bibnamefont{Lodish}},
  \bibinfo{author}{\bibfnamefont{A.}~\bibnamefont{Berk}},
  \bibinfo{author}{\bibfnamefont{P.}~\bibnamefont{Matsudaira}},
  \bibinfo{author}{\bibfnamefont{C.}~\bibnamefont{Kaiser}},
  \bibinfo{author}{\bibfnamefont{M.}~\bibnamefont{Krieger}},
  \bibinfo{author}{\bibfnamefont{M.}~\bibnamefont{Scott}},
  \bibinfo{author}{\bibfnamefont{S.}~\bibnamefont{Zipursky}}, \bibnamefont{and}
  \bibinfo{author}{\bibfnamefont{J.}~\bibnamefont{Darnell}},
  \emph{\bibinfo{title}{Molecular Cell Biology}} (\bibinfo{publisher}{W.H.
  Freeman and Company}, \bibinfo{year}{2003}).

\bibitem[{\citenamefont{Goldsbury et~al.}(2006)\citenamefont{Goldsbury, Mocanu,
  Thies, Kaether, Haass, Keller, Biernat, Mandelkow, and
  Mandelkow}}]{mandelkow5}
\bibinfo{author}{\bibfnamefont{C.}~\bibnamefont{Goldsbury}},
  \bibinfo{author}{\bibfnamefont{M.-M.} \bibnamefont{Mocanu}},
  \bibinfo{author}{\bibfnamefont{E.}~\bibnamefont{Thies}},
  \bibinfo{author}{\bibfnamefont{C.}~\bibnamefont{Kaether}},
  \bibinfo{author}{\bibfnamefont{C.}~\bibnamefont{Haass}},
  \bibinfo{author}{\bibfnamefont{P.}~\bibnamefont{Keller}},
  \bibinfo{author}{\bibfnamefont{J.}~\bibnamefont{Biernat}},
  \bibinfo{author}{\bibfnamefont{E.}~\bibnamefont{Mandelkow}},
  \bibnamefont{and} \bibinfo{author}{\bibfnamefont{E.-M.}
  \bibnamefont{Mandelkow}}, \bibinfo{journal}{Traffic}
  \textbf{\bibinfo{volume}{7}}, \bibinfo{pages}{1} (\bibinfo{year}{2006}).

\bibitem[{\citenamefont{Lipowsky et~al.}(2001)\citenamefont{Lipowsky, Klumpp,
  and Nieuwenhuizen}}]{LipowskyKN01}
\bibinfo{author}{\bibfnamefont{R.}~\bibnamefont{Lipowsky}},
  \bibinfo{author}{\bibfnamefont{S.}~\bibnamefont{Klumpp}}, \bibnamefont{and}
  \bibinfo{author}{\bibfnamefont{T.M.}~\bibnamefont{Nieuwenhuizen}},
  \bibinfo{journal}{Phys. Rev. Lett.} \textbf{\bibinfo{volume}{87}},
  \bibinfo{pages}{108101} (\bibinfo{year}{2001}).

\bibitem[{\citenamefont{Lipowsky and Klumpp}(2005)}]{LipowskyK05}
\bibinfo{author}{\bibfnamefont{R.}~\bibnamefont{Lipowsky}} \bibnamefont{and}
  \bibinfo{author}{\bibfnamefont{S.}~\bibnamefont{Klumpp}},
  \bibinfo{journal}{Physica A} \textbf{\bibinfo{volume}{352}},
  \bibinfo{pages}{53} (\bibinfo{year}{2005}).

\bibitem[{\citenamefont{Parmeggiani et~al.}(2003)\citenamefont{Parmeggiani,
  Franosch, and Frey}}]{pff1}
\bibinfo{author}{\bibfnamefont{A.}~\bibnamefont{Parmeggiani}},
  \bibinfo{author}{\bibfnamefont{T.}~\bibnamefont{Franosch}}, \bibnamefont{and}
  \bibinfo{author}{\bibfnamefont{E.}~\bibnamefont{Frey}},
  \bibinfo{journal}{Phys. Rev. Lett.} \textbf{\bibinfo{volume}{90}},
  \bibinfo{pages}{086601} (\bibinfo{year}{2003}).

\bibitem[{\citenamefont{Parmeggiani et~al.}(2004)\citenamefont{Parmeggiani,
  Franosch, and Frey}}]{pff2}
\bibinfo{author}{\bibfnamefont{A.}~\bibnamefont{Parmeggiani}},
  \bibinfo{author}{\bibfnamefont{T.}~\bibnamefont{Franosch}}, \bibnamefont{and}
  \bibinfo{author}{\bibfnamefont{E.}~\bibnamefont{Frey}},
  \bibinfo{journal}{Phys. Rev. E} \textbf{\bibinfo{volume}{70}},
  \bibinfo{pages}{046101} (\bibinfo{year}{2004}).

\bibitem[{\citenamefont{Popkov et~al.}(2003)\citenamefont{Popkov, Rakos,
  Willmann, Kolomeisky, and Sch\"utz}}]{popkov1}
\bibinfo{author}{\bibfnamefont{V.}~\bibnamefont{Popkov}},
  \bibinfo{author}{\bibfnamefont{A.}~\bibnamefont{Rakos}},
  \bibinfo{author}{\bibfnamefont{R.~D.} \bibnamefont{Willmann}},
  \bibinfo{author}{\bibfnamefont{A.~B.} \bibnamefont{Kolomeisky}},
  \bibnamefont{and} \bibinfo{author}{\bibfnamefont{G.~M.}
  \bibnamefont{Sch\"utz}}, \bibinfo{journal}{Phys. Rev. E}
  \textbf{\bibinfo{volume}{67}}, \bibinfo{pages}{066117}
  (\bibinfo{year}{2003}).

\bibitem[{\citenamefont{R\'akos et~al.}(2003)\citenamefont{R\'akos, Paessens,
  and Sch\"utz}}]{schuetz_nonergodic}
\bibinfo{author}{\bibfnamefont{A.}~\bibnamefont{R\'akos}},
  \bibinfo{author}{\bibfnamefont{M.}~\bibnamefont{Paessens}}, \bibnamefont{and}
  \bibinfo{author}{\bibfnamefont{G.~M.} \bibnamefont{Sch\"utz}},
  \bibinfo{journal}{Phys. Rev. Lett.} \textbf{\bibinfo{volume}{91}},
  \bibinfo{pages}{238302} (\bibinfo{year}{2003}).

\bibitem[{\citenamefont{Greulich}(2006)}]{diplom}
\bibinfo{author}{\bibfnamefont{P.}~\bibnamefont{Greulich}},
  \bibinfo{type}{Diploma thesis}, \bibinfo{school}{Universität zu Köln}
  (\bibinfo{year}{2006}).

\bibitem[{\citenamefont{Pierobon et~al.}(2006)\citenamefont{Pierobon, Mobilia,
  Kouyos, and Frey}}]{pff_1def}
\bibinfo{author}{\bibfnamefont{P.}~\bibnamefont{Pierobon}},
  \bibinfo{author}{\bibfnamefont{M.}~\bibnamefont{Mobilia}},
  \bibinfo{author}{\bibfnamefont{R.}~\bibnamefont{Kouyos}}, \bibnamefont{and}
  \bibinfo{author}{\bibfnamefont{E.}~\bibnamefont{Frey}},
  \bibinfo{journal}{Phys. Rev. E} \textbf{\bibinfo{volume}{74}},
  \bibinfo{pages}{031906} (\bibinfo{year}{2006}).

\bibitem[{\citenamefont{Evans et~al.}(2003)\citenamefont{Evans, Juhasz, and
  Santen}}]{evans_pff}
\bibinfo{author}{\bibfnamefont{M.~R.} \bibnamefont{Evans}},
  \bibinfo{author}{\bibfnamefont{R.}~\bibnamefont{Juhasz}}, \bibnamefont{and}
  \bibinfo{author}{\bibfnamefont{L.}~\bibnamefont{Santen}},
  \bibinfo{journal}{Phys. Rev. E} \textbf{\bibinfo{volume}{68}},
  \bibinfo{pages}{026117} (\bibinfo{year}{2003}).

\bibitem[{\citenamefont{Janowsky and Lebowitz}(1992)}]{lebowitz1}
\bibinfo{author}{\bibfnamefont{S.A.}~\bibnamefont{Janowsky}} \bibnamefont{and}
  \bibinfo{author}{\bibfnamefont{J.L.}~\bibnamefont{Lebowitz}},
  \bibinfo{journal}{Phys. Rev. A} \textbf{\bibinfo{volume}{45}},
  \bibinfo{pages}{618} (\bibinfo{year}{1992}).

\bibitem[{\citenamefont{Chou and Lakatos}(2004{\natexlab{b}})}]{bn_quasiexact}
\bibinfo{author}{\bibfnamefont{T.}~\bibnamefont{Chou}} \bibnamefont{and}
  \bibinfo{author}{\bibfnamefont{G.}~\bibnamefont{Lakatos}},
  \bibinfo{journal}{Phys. Rev. Lett.} \textbf{\bibinfo{volume}{93}},
  \bibinfo{pages}{198101} (\bibinfo{year}{2004}{\natexlab{b}}).

\bibitem[{\citenamefont{Kolomeisky et~al.}(1998)\citenamefont{Kolomeisky,
  Schütz, Kolomeisky, and Straley}}]{kolomeisky_shockdyn}
\bibinfo{author}{\bibfnamefont{A.}~\bibnamefont{Kolomeisky}},
  \bibinfo{author}{\bibfnamefont{G.}~\bibnamefont{Schütz}},
  \bibinfo{author}{\bibfnamefont{E.}~\bibnamefont{Kolomeisky}},
  \bibnamefont{and} \bibinfo{author}{\bibfnamefont{J.}~\bibnamefont{Straley}},
  \bibinfo{journal}{J. Phys. A} \textbf{\bibinfo{volume}{31}},
  \bibinfo{pages}{6911} (\bibinfo{year}{1998}).

\bibitem[{\citenamefont{Sch\"utz}(2001)}]{shockdyn1}
\bibinfo{author}{\bibfnamefont{G.}~\bibnamefont{Sch\"utz}}, in
  \emph{\bibinfo{booktitle}{Phase Transitions and Critical Phenomena}}, edited
  by \bibinfo{editor}{\bibfnamefont{C.}~\bibnamefont{Domb}} \bibnamefont{and}
  \bibinfo{editor}{\bibfnamefont{J.}~\bibnamefont{Lebowitz}}
  (\bibinfo{publisher}{Academic Press}, \bibinfo{year}{2001}),
  vol.~\bibinfo{volume}{19}.

\bibitem[{\citenamefont{Popkov and Schütz}(1999)}]{popkov2}
\bibinfo{author}{\bibfnamefont{V.}~\bibnamefont{Popkov}} \bibnamefont{and}
  \bibinfo{author}{\bibfnamefont{G.~M.} \bibnamefont{Schütz}},
  \bibinfo{journal}{Europhys. Lett.} \textbf{\bibinfo{volume}{48}},
  \bibinfo{pages}{257} (\bibinfo{year}{1999}).

\bibitem[{\citenamefont{Greulich et~al.}(2007)\citenamefont{Greulich, Garai,
  Nishinari, Schadschneider, and Chowdhury}}]{kif1a2}
\bibinfo{author}{\bibfnamefont{P.}~\bibnamefont{Greulich}},
  \bibinfo{author}{\bibfnamefont{A.}~\bibnamefont{Garai}},
  \bibinfo{author}{\bibfnamefont{K.}~\bibnamefont{Nishinari}},
  \bibinfo{author}{\bibfnamefont{A.}~\bibnamefont{Schadschneider}},
  \bibnamefont{and}
  \bibinfo{author}{\bibfnamefont{D.}~\bibnamefont{Chowdhury}},
  \bibinfo{journal}{Phys. Rev. E} \textbf{\bibinfo{volume}{75}},
  \bibinfo{pages}{041905} (\bibinfo{year}{2007}).

\bibitem[{\citenamefont{Nishinari et~al.}(2005)\citenamefont{Nishinari, Okada,
  Schadschneider, and Chowdhury}}]{kif1a1}
\bibinfo{author}{\bibfnamefont{K.}~\bibnamefont{Nishinari}},
  \bibinfo{author}{\bibfnamefont{Y.}~\bibnamefont{Okada}},
  \bibinfo{author}{\bibfnamefont{A.}~\bibnamefont{Schadschneider}},
  \bibnamefont{and}
  \bibinfo{author}{\bibfnamefont{D.}~\bibnamefont{Chowdhury}},
  \bibinfo{journal}{Phys Rev Lett} \textbf{\bibinfo{volume}{95}},
  \bibinfo{pages}{118101} (\bibinfo{year}{2005}).

\bibitem[{\citenamefont{Chowdhury et~al.}(2000)\citenamefont{Chowdhury, Santen,
  and Schadschneider}}]{traffic1}
\bibinfo{author}{\bibfnamefont{D.}~\bibnamefont{Chowdhury}},
  \bibinfo{author}{\bibfnamefont{L.}~\bibnamefont{Santen}}, \bibnamefont{and}
  \bibinfo{author}{\bibfnamefont{A.}~\bibnamefont{Schadschneider}},
  \bibinfo{journal}{Phys. Rep.} \textbf{\bibinfo{volume}{329}},
  \bibinfo{pages}{199} (\bibinfo{year}{2000}).

\bibitem[{\citenamefont{Boldrighini et~al.}(1989)\citenamefont{Boldrighini,
  Cosimi, Frigio, and Nunes}}]{scp1}
\bibinfo{author}{\bibfnamefont{C.}~\bibnamefont{Boldrighini}},
  \bibinfo{author}{\bibfnamefont{G.}~\bibnamefont{Cosimi}},
  \bibinfo{author}{\bibfnamefont{S.}~\bibnamefont{Frigio}}, \bibnamefont{and}
  \bibinfo{author}{\bibfnamefont{M.~G.} \bibnamefont{Nunes}},
  \bibinfo{journal}{J. Stat. Phys.} \textbf{\bibinfo{volume}{55}},
  \bibinfo{pages}{611} (\bibinfo{year}{1989}).

\end{thebibliography}
\end{document}